\documentclass[11pt]{article}
\pdfoutput=1  
\usepackage{jcappubmod}

\usepackage{graphicx}
\usepackage{color}
\usepackage{amssymb}
\usepackage{amsmath}
\usepackage{mathabx}
\usepackage{stmaryrd}
\usepackage{multirow}
\usepackage{multicol}
\usepackage{amscd}
\usepackage{mfpic}
\usepackage{verbatim}
\usepackage{hyperref}
\usepackage{pstricks,pst-node,pst-text,pst-3d}
\usepackage{ftnxtra}
\usepackage{fnpos} 

\definecolor{cream}{rgb}{.97, .95, .88}
\definecolor{darkcream}{rgb}{1., .88, .5}
\definecolor{lightpink}{rgb}{0.98, 0.88, 0.87}
\definecolor{lightwhite}{rgb}{1., 0.98, 0.95}
\definecolor{lightsalmon}{rgb}{1., 0.95, 0.90}
\definecolor{lightviolet}{rgb}{0.9, 0.8, 0.9}
\definecolor{lightgray}{rgb}{.96, .96, .96}  
\definecolor{lgray}{rgb}{.75, .75, .75}
\definecolor{LemonChiffon}{rgb}{0.95, 1., 0.7}
\definecolor{lightolivegreen}{rgb}{0.84, 0.89, 0.25}
\definecolor{lightgreen}{rgb}{.664, 1., .52}
\definecolor{llgreen}{rgb}{.900, .983, .960}
\definecolor{tristle}{rgb}{0.87, 0.67, 0.77} 
\definecolor{pink}{rgb}{0.95, 0.45, 0.75}
\definecolor{magenta}{rgb}{1., 0, 1.}
\definecolor{violet}{rgb}{0.9, 0.20, 0.85}
\definecolor{darkolivegreen}{rgb}{0.55, 0.65, 0.35}
\definecolor{maroon}{rgb}{0.7, 0.26, 0.56}
\definecolor{lightmaroon}{rgb}{0.85, 0.38, 0.58}
\definecolor{darkmaroon}{rgb}{0.604, 0.169, 0.451}
\definecolor{ddarkmaroon}{rgb}{0.2, 0.03125, 0.150}
\definecolor{mediumorchid}{rgb}{0.8, 0.33, 0.83}
\definecolor{mediumorchidd}{rgb}{1., 0.33, 0.63}
\definecolor{darkgreen}{rgb}{0.1, 0.6, 0.13}
\definecolor{lightyellow}{rgb}{1., 1., 0.82}
\definecolor{turquoise}{rgb}{0.042, 0.586, 0.512}
\definecolor{turquoisel}{rgb}{0.66, 0.94, 0.83}
\definecolor{darkturquoise}{rgb}{0.21, 0.55, 0.50}
\definecolor{coral}{rgb}{1., 0.6, 0.21}
\definecolor{lightorange}{rgb}{1., 0.88, 0.75}
\definecolor{orangered}{rgb}{1., 0.5, 0.}
\definecolor{orange}{rgb}{1., 0.65, 0.1}
\definecolor{orangel}{rgb}{1., .85, .3}
\definecolor{darkorange}{rgb}{0.875, 0.4, 0.204}
\definecolor{ddarkorange}{rgb}{.675, .218, .05}
\definecolor{bluesky}{rgb}{0.48, 0.53, 1.}
\definecolor{gold}{rgb}{1., 0.85, 0.25}
\definecolor{goldd}{rgb}{0.95, 0.75, 0.05}
\definecolor{darkviolet}{rgb}{0.54, 0.04, 0.84}
\definecolor{ddarkviolet}{rgb}{.382, .063, .657}
\definecolor{lightblue}{rgb}{0.30, 0.86, 0.89}
\definecolor{LightBlue}{rgb}{0.68, 0.85, 0.9}
\definecolor{lblue}{rgb}{0.78, 0.90, 0.95}
\definecolor{darkblue}{rgb}{.105, .308, .707}
\definecolor{lightmaroon}{rgb}{0.85, 0.38, 0.58}
\definecolor{darkmaroon}{rgb}{0.604, 0.169, 0.451}
\definecolor{darkpink}{rgb}{0.879, 0.020, 0.766}
\definecolor{ddarkpink}{rgb}{0.738, 0.195, 0.406}
\definecolor{grey}{rgb}{0.717, 0.717, 0.717}
\definecolor{lightgrey}{rgb}{0.800, 0.800, 0.800}
\definecolor{brown}{rgb}{0.740, 0.323, 0.182}
\definecolor{redbrown}{rgb}{.575, .158, .05}
\definecolor{darkbrown}{rgb}{0.34, 0.25, 0.05}
\definecolor{orangebrown}{rgb}{0.433, 0.262, 0.06}
\definecolor{pinkl}{rgb}{1., 0.788, 0.918}
\definecolor{salmon}{rgb}{1., 0.66, 0.5}
\definecolor{lightbrown}{rgb}{0.703, 0.508, 0.121}

\def\etal{{\it et al.}}

\def\Name#1#2 {{#2} {#1, }}

\def\Journal#1#2#3#4{{#1} {\bf #2}, (#3) #4}
\def\cir#1{{\GCN} #1}

\def\AA{\em A.\& A.}

\def\APJ{\em ApJ.}
\def\APJL{\em ApJ.Lett.}

\def\AST{\em Astron. J.}

\def\CQG{\em Class.Quant.Grav.}
\def\EAS{\em EAS Publications Series}

\def\GCN{\em GCN Circ.}

\def\IMD{{\em Int. J. Mod. Phys.} D}

\def\JCA{\em J. Cosmol. Astrop. Phys.}

\def\MRA{\em MNRAS}

\def\NAT{\em Nature}
\def\NATA{\em Nature Astro.}

\def\PRD{{\em Phys. Rev.} D}

\def\SCI{\em Science}

\def\SSC{\em Space Sci.}

\def\be{\begin{equation}}
\def\ee{\end{equation}}
\def\bea{\begin{eqnarray}}
\def\eea{\end{eqnarray}}
\def\bes{\begin{equation*}}
\def\ees{\end{equation*}}
\def\beas{\begin{eqnarray*}}
\def\eeas{\end{eqnarray*}}

\title{Late afterglows of GW/GRB 170817A}

\author[a,b]{Houri~Ziaeepour}
\affiliation [a]{Institut UTINAM, CNRS UMR 6213, Observatoire de Besan\c{c}on, Universit\'e de Franche Compt\'e, 41 bis ave. de l'Observatoire, BP 1615, 25010 Besan\c{c}on, France}
\affiliation [b]{Mullard Space Science Laboratory, University College London, Holmbury St. Mary, GU5 6NT, Dorking, UK}
\emailAdd{houriziaeepour@gmail.com}

\begin{document}



\abstract
{The gamma-ray burst that followed the first detection of gravitational waves from the merger 
of a Binary Neutron Stars (BNS) and its low energy counterparts were in many respects unusual 
and challenge our understanding of mechanisms involved in their production. In a previous work 
we used a phenomenological formulation of relativistic shocks and synchrotron emission to analyse 
the prompt gamma-ray emission of GW/GRB 170817A. Here we use the same model to analyse late 
afterglows of this event. The main goal is to see whether synchrotron emission alone can explain 
the late afterglows. We find that collision between a mildly relativistic outflow from the merger 
with a Lorentz factor of $\sim 1.2-3$ and the ISM/circumburst material can explain observations, 
if the synchrotron self-absorption of radio emission and local extinction of optical/IR photons 
are taken into account. In absence of a significant extinction, an additional source of X-ray is 
necessary to explain the data. These conclusions are in large extend independent of the model 
used here and can be deduced directly from data. We also show that at the time of its encounter 
with circumburst material the outflow could have been still mildly magnetized. The origin for 
optical extinction could be a dust rich old faint star cluster surrounding the BNS, which 
additionally had helped its formation and merger. Such an environment evades present observational 
constraints and is consistent with our conclusions about properties and evolution of the progenitor 
neutron stars obtained from analysis of the prompt gamma-ray emission. Alternatively, if the 
synchrotron emission was produced internally through collisions of density shells, the extinction 
might have occurred inside the outflow itself rather than externally. The most plausible additional 
source of X-ray is the decay of medium and heavy isotopes produced by the kilonova, including 
r-processes, and the recombination of cooled electrons. The contribution of these processes should 
be quantified in future works.}


\maketitle

\section{Introduction} \label{sec:intro}
The afterglow of GRB 170817A associated to the Gravitational Wave (GW) event GW 170817 is the only 
short GRB with very long follow up in a broad band of energies, from X-ray to radio. The motivation 
for studying these faint emissions is not only investigating the fate of a Binary Neutron Star 
(BNS) merger remnant, but also understanding the origin of unusual faintness of the prompt gamma-ray 
emission of this event and peculiar behaviour of its afterglows, namely their early faintness 
and late brightening\footnote{In this work by {\it early afterglows} we mean emissions in X-ray and 
lower energy bands from $\gtrsim 70$~sec - the average slew delay of the Neil Gehrels Swift 
Observatory~\citep{swift} - up to $\mathcal{O}(1) \times 10^5$~sec, i.e. a couple of days after 
gamma-ray trigger. This is the time interval in which the afterglow of most short GRBs detected by 
the Swift have been observed. Usually no later attempt is made to detect afterglows of GRBs without 
an X-ray counterpart in this interval and GRB 170817A is an exception.}. Explanations suggested in 
the literature include: off axis observation of both prompt and afterglow of an otherwise ordinary 
short GRB~\citep{grboffaxprobab,gw170817latexraystructjet,gw170817xray,gw170817latebroad,grboffaxprobab0}; 
prompt gamma-ray emission from breakout of a chocked jet and afterglow from a mildly 
relativistic\footnote{In the literature the range of Lorentz factor called 
{\it mildly relativistic} varies and depends on the context and author. Here we use this term for 
fluids with a Lorentz factor in the range of $1.2 - 5$, corresponding to velocity range 
$0.5c - 0.98c$, where $c$ is the speed of light.} structured jet~\citep{structuredjet,gw170817lateopt} 
or a cocoon~\citep{cocoon,grbthermalcocoon,grbcocoon,gw170817cocoon0,gw170817cocoon,gw170817lateradio,gw170817cocoonsimul}; 
prompt emission from a structured jet and afterglow from a roughly spherical 
outflow~\citep{gw170817model,gw170817latexary}; and variation of these models: jet breakout from 
wind~\citep{gw170817earlyradio0}, jet-merger-ejecta interaction~\citep{jetmergerejectint}, 
neutron-rich isotropic fireball~\citep{gw170817fireball}, and gamma-ray emission from a wide 
spiraling outflow~\citep{gw170817promptdisk}.

Predictions of these models have been mostly compared only with late X-ray and radio afterglows, 
because no observation in these bands was performed before $\sim T+1.6$~days, where $T$ is the 
Fermi-GBM~\citep{fermi} trigger time. No counterpart was detected before $~\sim T+10$~days and 
$~\sim T+16$~days in X-ray and radio wavelengths, respectively. The earliest observations in 
optical/IR began at the same time as in X-ray and radio, and leaded to the detection of a 
counterpart. However, emissions in optical/IR bands are believed to be dominated by a kilonova 
originated from the ejection of hot low velocity material from the disk/torus of the 
merger~\citep{gw170817optkilonova,grb130603bkilonova,gw170817bluekilonova,gw170817bluekilonovapol,gw170817bluekilonovamod,gw170817optkilonovath,gw170817kilonovaspeed,gw170817kilonovadetail} 
and not from the GRB producing jet. Nonetheless, a contribution from the polar outflow, which 
generates a relativistic jet and a GRB may be necessary to explain these 
observations~\citep{gw170817optkilonova,gw170817bluekilonova}. We remind that most works on the 
unusual properties of GRB 170817A do not compare their models with the prompt emission data obtained 
by the Fermi-GRB~\citep{gw170817fermi} and the Integral-IBIS~\citep{gw170817integral} instruments, 
and only their consistency with the data is discussed. 

In a previous work~\citep{hourigw170817} we used a phenomenological formulation of relativistic 
shocks and their synchrotron/self-Compton emission~\citep{hourigrb,hourigrbmag} to model in detail 
light curves and spectrum of the prompt gamma-ray emission of GRB 170817A. It shows that the 
most plausible Lorentz factor for the relativistic jet of this burst is $\Gamma\sim 100$. However, 
parameters of the model are degenerate and a low Lorentz factor of $\sim 10$ also fits the data, 
if the energy transfer to accelerated electron is very efficient and close to highest values seen in 
Particle In Cell (PIC) simulations~\citep{fermiaccspec}. In any case, these Lorentz factors are in 
the highest limit of values suggested in the literature, but at least few folds less than what is 
predicted for more typical short GRBs, for instance GRB 130603B~\citep{grb130603b} which was also 
associated to a kilonova~\citep{grb130603bkilonova,grb130603bkilonova0}. Moreover, it is found that 
density and extent of the relativistic jet of GRB 170817A were more than an order of magnitude less 
than GRB 130603B. Using the results of BNS merger simulations, we concluded that plausible reasons 
behind the weak jet in GW 170817 BNS merger event were: the old age of neutron star progenitors 
and their reduced magnetic field and consequently availability of less energy for formation and 
acceleration of the jet; the closeness of progenitors masses; and finally the evolutionary history 
of the BNS and perturbation of its orbit and spin of neutron stars due to encounter with other stars 
and gravitational disturbances, probably in the dense environment of a star cluster. These event 
misaligned spins of neutron stars and along with reduced magnetic fields of stars leaded to a 
weaker than usual polar dynamical ejecta during merger.

To complete our analysis of GW/GRB 170817A electromagnetic data, in this work we present simulations 
of late afterglows according to the same phenomenological model as the one used for the modelling 
of prompt emission. Several groups have used approximate analytical expressions for synchrotron 
emission from shocks to fit light curves of afterglows of 
GW/GRB 170817A~\citep{gw170817latebroad,gw170817lateradio,gw170817cocoonsimul,gw170817latefasttail}, 
and obtain good fit to the data. In this approach models used for fitting the data depend on a few 
parameters and it is not difficult to obtain an acceptable fit by adjusting them. 
Even in works based on MHD simulations such as~\citep{gw170817latefasttail}, the MHD is used to 
predict properties of the outflow such as its kinetic and density rather than the observed 
synchrotron emission. Therefore, a goal of present work is to see whether simulation of shocks and 
synchrotron emission using physically motivated expressions for fundamental quantities such as 
densities, magnetic field, etc. confirm the prior assumption of synchrotron emission as origin of 
afterglows, or contribution from other processes would be needed to explain the data. Considering 
the complexity of the physics of BNS merger and its remnants, it is expected that after weakening 
of shocks in the ejecta, subdominant processes such as decay of long living isotopes in the 
kilonova remnant become detectable.

We briefly review the model and its parameters in Sec. \ref{sec:model}. Methodology of modelling, 
initial conditions, and assumptions are discussed in Sec. \ref{sec:agsimul}. Results of the 
simulations are presented in Sec. \ref{sec:simul}. Interpretation of the results are discussed in 
Sec. \ref {sec:discuss}, and outlines in Sec. \ref {sec:outline}. In Appendix \ref{app:abs} we 
calculate synchrotron self-absorption index for the phenomenological shock model 
of~\citep{hourigrb,hourigrbmag}. Appendix \ref{app:modes} summarizes phenomenological expressions 
used for modelling the width of synchrotron emitting region in the model. 

\section{Model} \label{sec:model}
The phenomenological model of~\citep{hourigrb,hourigrbmag} assumes that GRB emissions are produced 
by synchrotron/self-Compton processes in a dynamically active region in the head front of shocks 
between density shells inside a relativistic cylindrical jet for prompt and with surrounding 
material for afterglows in lower energies. In addition to the magnetic field generated by Fermi 
processes in the active region an external magnetic field precessing with respect to the jet 
axis may contribute in the production of synchrotron emission. An essential aspect of this model, 
which distinguishes it from other phenomenological GRB formulations, is the evolution of parameters 
with time. Moreover, simulation of each burst consists of a few time intervals - {\it regimes} - 
each corresponding to an evolution rule (model) for phenomenological quantities such as: 
fraction of kinetic energy transferred to fields and its variation; variation of the thickness of 
synchrotron/self-Compton emitting {\it active} region; etc. Division of simulated bursts to these 
intervals allows to change parameters and phenomenological evolution rules which are kept constant 
during one time interval. Physical motivation for such fine-tuning is the fact that GRB producing 
shocks are in a highly non-equilibrium and fast varying state. In fact, multiple variation of the 
slope of afterglows light curves, which presumably are produced by external shocks on the ISM or 
circumburst material is an evidence that they are not completely uniform and their anisotropies 
affect the emission.

This simplified formulation of shocks is originally constructed for prompt and early afterglow 
emissions, which presumably are generated by a fast and compact ejecta. The main approximation in 
the model is the assumption that properties of matter in colliding shells is uniform and 
self-similar. According to this approximation the evolution of shock and its synchrotron emission 
depends only on time or equivalently distance from the center, rather than to both time and 
distance as independent variables. However, as suggested in the 
literature~\citep{gw170817cocoon,gw170817cocoon0,gw170817lateradio,gw170817latefasttail,gw170817latedecline} 
and we give more arguments in its favour in the next section, the observed late afterglows of 
GW/GRB 170817A might have been produced by a continuous flow with varying spatiotemporal 
characteristics. To take into account slow variation of characteristics of the outflow, we simulate 
the model with different densities and speeds, presenting slow variation of physical properties.
Each simulation can be considered as presenting the outflow at different epochs or fraction of 
material with corresponding properties. If the underlying priors are correct, a linear 
interpolation between these models should be a good approximation of emission from a slowly 
varying or nonuniform outflow. 


Table \ref{tab:paramdef} summarizes parameters of the model. Despite their long list, simulation 
of typical long and short GRBs in~\citep{hourigrbmag} shows that the range of values which lead 
to realistic bursts are fairly restricted. Evidently, the range of some parameters for simulation 
of prompt and afterglow are very different. In particular, we assume that the ISM or surrounding 
material is at rest with respect to a far observer at the same redshift. They are non-relativistic 
and in comparison with relativistic outflows their velocity dispersion is negligible. Therefore, 
for external shocks the Lorentz factor of slow shell $\Gamma$ is always 1.

\begin{table}
\begin{center}
\caption{Parameters of the phenomenological relativistic shock model \label{tab:paramdef}}
\vspace{0.5cm}
\begin{tabular}{| p{2.5cm} | p{12cm} |}
\hline
Model (mod.) & Model for evolution of active region with distance from central engine; See 
Appendix \ref{app:modes} and~\citep{hourigrb,hourigrbmag} for more details. \\
$r_0$ (cm) & Initial distance of shock front from central engine. \\
$\Delta r_0$ & Initial (or final, depending on the model) thickness of active region. \\
$p$ & Slope of power-law spectrum for accelerated electrons; See eq. (3.8) of~\citep{hourigrbmag}. \\
$p_1,~p_2$ & Slopes of double power-law spectrum for accelerated electrons; See eq. (3.14) 
of~\citep{hourigrbmag}. \\
$\gamma_{cut}$ & Cut-off Lorentz factor in power-law with exponential cutoff spectrum for 
accelerated electrons; See eq. (3.11) of~\citep{hourigrbmag}. \\
$\gamma'_0$ & Initial Lorentz factor of fast shell with respect to slow shell. \\
$\delta$ & Index in the model defined in eq. (3.29) of~\citep{hourigrbmag}. \\
$Y_e$ & Electron yield defined as the ratio of electron (or proton) number density to baryon number 
density. \\
$\epsilon_e$ & Fraction of the kinetic energy of falling baryons of fast shell transferred to 
leptons in the slow shell (defined in the slow shell frame). \\
$\alpha_e$ & Power index of $\epsilon_e$ as a function of $r$. \\
$\epsilon_B$ & Fraction of baryons kinetic energy transferred to induced magnetic field in the 
active region. \\
$\alpha_B$ & Power index of $\epsilon_B$ as a function of $r$. \\
$N'$ & Baryon number density of slow shell. \\
$\kappa$ & Power-law index for N' dependence on $r'$. \\
$n'_c$ & Column density of fast shell at $r'_0$. \\
$\Gamma$ & Lorentz factor of slow shell with respect to far observer.\\
\hline
\end{tabular}
\end{center}
{\small
\begin{description}
\item{$\star$} The phenomenological model discussed in~\citep{hourigrb} and its 
simulation~\citep{hourigrbmag} depends only on the combination $Y_e\epsilon_e$. For this reason 
only the value of this combination is given for simulations.
\item{$\star$} The model neglects variation of physical properties along the jet or active region. 
They only depend on the average distance from center $r$, that is $r-r_0 \propto t-t_0$.
\item{$\star$} Quantities with prime are defined with respect to rest frame of slow shell, and 
without prime with respect to central object, which is assumed to be at rest with respect to 
a far observer. Power indices do not follow this rule.
\end{description}
}
\end{table}

\section{Late afterglow of GW/GRB 170817A} \label{sec:agsimul}
It is unlikely that the observed electromagnetic emissions from GW/GRB 170817A event at 
$\gtrsim T+10$~days can be due to the remnant of the relativistic jet which generated the observed 
prompt gamma-ray. The early afterglows of short GRBs, which are presumably produced by weaker 
internal and external shocks of the relativistic jet~\citep{xrtafterglow}, are usually observable 
for few days, because their flux declines very quickly in all bands~\citep{grbshortag}. Therefore, 
as other authors concluded~\citep{gw170817lateradio,gw170817latexary,gw170817latedecline}, at 
$\gtrsim T+10$~days the relativistic jet of GRB 170817A, which according to our analysis its kinetic 
energy, extent, and density were much smaller than typical short GRBs~\citep{hourigw170817}, has 
been weakened and dissipated by interaction with its surrounding material~\citep{hourigrbmag} and 
could not have significant contribution in the observed emissions. In this case, it would be 
indistinguishable from the slower part of dynamical ejecta.

The observed late brightening of GRB 170817A and a few other 
short~\citep{grb130603bxray,grb091109b} and long~\citep{grb060712latex,grb060807latex} GRBs should 
be due to the emission from other components of ejecta and/or other processes. Candidates suggested 
in the literature are: MHD instabilities leading to increase in magnetic energy 
dissipation~\citep{grbjetsimulinstab,grbjetsimulinstab0}; external shocks generated by the collision 
between the ISM or circumburst material and a mildly relativistic thermal cocoon ejected at the 
same time as the ultra-relativistic component and lagged with respect to the 
jet~\citep{nakarpoynting,cocoon,gw170817lateopt,gw170817latefasttail}; late outflows from a long 
lasting accretion disk~\citep{latexrayexcess}; fall-back of ejected matter to the central black 
hole~\citep{gw170817agfallback}. 

The off-axis view of a structured jet predicts late brightening of 
afterglows~\citep{grboffaxis,gw170817latexraystructjet,gw170817latexary,grboffaxprobab}. However, 
the observed decline of X-ray flux at $\lesssim T+134$ days~\citep{gw170817latedecline} is 
inconsistent with simulations of significantly off-axis emission~\citep{gw170817latexraystructjet}, 
which predict a break after a few hundred days. Other simulations, for instance those reported 
by~\citep{gw170817latexary,gw170817offaxecocoondisc} predict earlier break, but cannot discriminate 
between off-axis and cocoon (structured jet) models~\citep{gw170817lateradio} and need polarimetry 
and imaging to discriminate between them~\citep{gw170817offaxecocoondisc}. 

Our prior assumption is that late afterglows of GRB 17081A and their brightening is due to 
synchrotron emission from the mildly relativistic remnant of dynamical ejecta during BNS merger, 
which from now on we simply call {\it the outflow}. Indeed, General Relativistic 
Magneto-Hydro-Dynamic (GRMHD) simulations of BNS merger show poleward mildly relativistic - with a 
Lorentz factor of $\lesssim 4$ - mass 
ejection~\citep{nstarbhmergsimul2,nsmergerrprocsimulout,nsmergerrprocsimulhres}. 
Only a small fraction of this ejecta is accelerated to ultra-relativistic velocities by the transfer 
of Poynting energy of the ejecta to kinetic 
energy~\citep{grbjetsimul,grbjetsimul1,grbjetsimulinstab0}. 
The remaining material continues its trajectory and eventually collides with the ISM and/or 
circum-BNS material at a later time. If this scenario is correct, brightening of afterglows in 
other short GRBs is missed because observations were not sufficiently long. 

GRMHD simulations show that the opening angle of this outflow is $\lesssim 30^\circ$. The orbit 
inclination of the BNS merger associated to GW 170817 is estimated as 
$\theta_w \lesssim 18^\circ - 27^\circ$~\citep{gw170817decline}. Therefore, the viewing angle is 
expected to pass through the outflow and even if the latter have a somehow non-uniform velocity 
profile~\citep{gw170817lateopt}, in comparison with other uncertainties of the model its effect would 
be negligible for analysing the data. Indeed, the negligence of the inclination of the line of sight 
with respect to the outflows symmetry axis induces an error $\propto (1-\cos \theta_w) \lesssim 8\%$ 
for $\theta_w \approx 20^\circ$ in the simulated light curves, which is much less than other 
uncertainties of the model. For this reason, we will not discuss the issue of viewing angle further. 
Nonetheless, due to these simplifications and degeneracies between parameters of the model discussed 
in details in~\citep{hourigw170817}, the estimation of characteristics of the outflow by simulations 
presented in the next section should be considered as {\it order of magnitude} rather {\it exact}.

Although observations of GRBs afterglows are consistent with synchrotron/self-Compton mechanism 
as the main emission process, subdominant processes may become important and observable when the 
main source weakens. In particular, following the observation of a 
kilonova~\cite{gw170817optsss1,gw170817optkilonovath,gw170817optkilonova,gw170817bluekilonovamod} in 
GW/GRB 170817A event, we know that the decay of isotopes produced during merger and recombination 
of cooled electrons contribute to the emissions.

\subsection{Initial conditions} \label{sec:cocoon}
Before discussing how we have estimated initial conditions for the simulation of late afterglows 
of GRB 170817A we should remind that in the phenomenological model of~\citep{hourigrb,hourigrbmag} 
the jet is cylindrical and as long as the line of sight of the observer pass through it and its 
Lorentz factor is sufficiently high such that $\sin \theta_j > 1/\gamma'$, the effect of oblique 
view is negligible\footnote{For external shocks $\Gamma=1$. Thus, relative Lorentz factor of shells 
$\gamma'$ is equal to Lorentz factor of the jet $\Gamma_j$ with respect to a far observer at the same 
redshift as the source.}. Additionally, it is proved that for $\Gamma_j \gg 1$ the contribution 
of high latitude emission is small~\citep{hourigrb}. This condition is not fully satisfied by the 
late outflow, which is expected to be only mildly relativistic. Nonetheless, even for an outflow with a 
Lorentz factor $\Gamma_c \sim 2$ the value of $1/\Gamma^2_c \sim 1/4 \sim 1/2\Gamma_c$. Considering 
other theoretical and observational uncertainties in the modelling of the data, this amount of 
error should be tolerable when the aim is order of magnitude estimation of physical quantities 
which characterize the outflow and material surrounding the BNS merger.

Initial values of some parameters in Table \ref{tab:param} cannot be arbitrarily selected. For 
instant, the initial distance of the jet front from center when the presumed external shock begins 
must be consistent with the value of outflow's $\beta$ and the time of first detection of 
electromagnetic signal from outflow's external shock. For $\beta = 0.4~\text{and}~0.8$ corresponding 
to $\gamma' \approx 1.2~\text{and}~2.34$,\footnote{Because we assume that $\Gamma = 1$, i.e. the 
ISM/circumburst material is at rest with respect to a far observer at the same redshift, $\gamma'$ 
is the Lorentz factor of jet with respect to observer. Nonetheless, we keep $'$ for consistency of 
notation with Table \ref{tab:paramdef}.} the distance of the ejecta's front $r_0$ at the beginning 
of its collision with the ISM/surrounding material at $\lesssim 10$~days after merger must be 
$\approx 10^{16}~\text{cm and}~2 \times 10^{16}$~cm, respectively.

To estimate physically plausible column density of the outflow we use the results of GRMHD numerical 
simulations of BNS merger. The total amount of fast tail (dynamical) ejecta is expected to be 
$0.01-0.03 M_\odot$, where $M_\odot$ is the solar 
mass~\citep{gw170817optkilonova,gw170817rprocess,gw170817bluekilonova}. In the case of GW 170817 
event a larger ejecta of $m_{ejecta}\sim 0.03-0.05 M_\odot$ seems necessary to explain the 
bright UV/blue emission of kilonova at $\sim T+1.6$~days~\citep{gw170817bluekilonova}. However, a 
contribution from GRB 170817A afterglow in the early blue peak cannot be ruled 
out~\citep{gw170817optkilonova}, see also the discussion about the early afterglow of this burst 
in~\citep{hourigw170817}. We show later that under some conditions our simulations are 
consistent with the prediction of GRMHD simulations. 

Using the above estimated ejecta mass, the baryon number column density at distance 
$r_0 = c\beta t_0$ can be parametrized as: $n'_c = 0.3 \times 10^{25}\eta A / \beta^2 t^2_{day}$, where 
$\eta \equiv \pi/\theta_c$ and $A \equiv m_{ejecta}/ 0.01 M_\odot$. We use the approximate time of 
the first detection of the X-ray counterpart $t_0 \sim 10$~days to estimate order of magnitude 
value of column density $n'_c$ to be used in our simulations. For instance, for $\beta = 0.8$,  
$A = 1$, and a close to spherical outflow, we obtain $n'_c \sim 4.7 \times 10^{22}$~cm$^{-2}$ 
and $r_0 \sim 2 \times 10^{16}$~cm; for $\beta = 0.4$ the column density increases to 
$n'_c \sim 1.9 \times 10^{23}$~cm$^{-2}$ and initial distance decreases to $r_0 \sim 10^{16}$~cm. 
The values for distance to circumburst material found here correspond to typical distance of 
termination shock of ejecta from a star, that is where the ISM pressure becomes equal to wind pressure. 
In the cases of GW 170817 event circumburst material should consist of pre-merger ejecta - 
presumably the remnant of ejected matter during and after the formation of neutron stars and BNS  
and/or accreted material from environment. 

We should remind that due to the large parameter space of the model it was not possible to perform a 
systematic search for the best combination of parameters. This fact must be taken into account when 
simulations are compared with data and the values of parameters given in Table \ref{tab:param} 
should be considered as order of magnitude estimations.

\section{Simulations} \label{sec:simul}
\begin{figure}
\begin{center}
\begin{tabular}{p{7cm}p{7cm}p{7cm}}
Simul. 1 & \hspace{-1.5cm} Simul. 2 & \hspace{-3cm} Simul. 3 \\
\includegraphics[width=7cm]{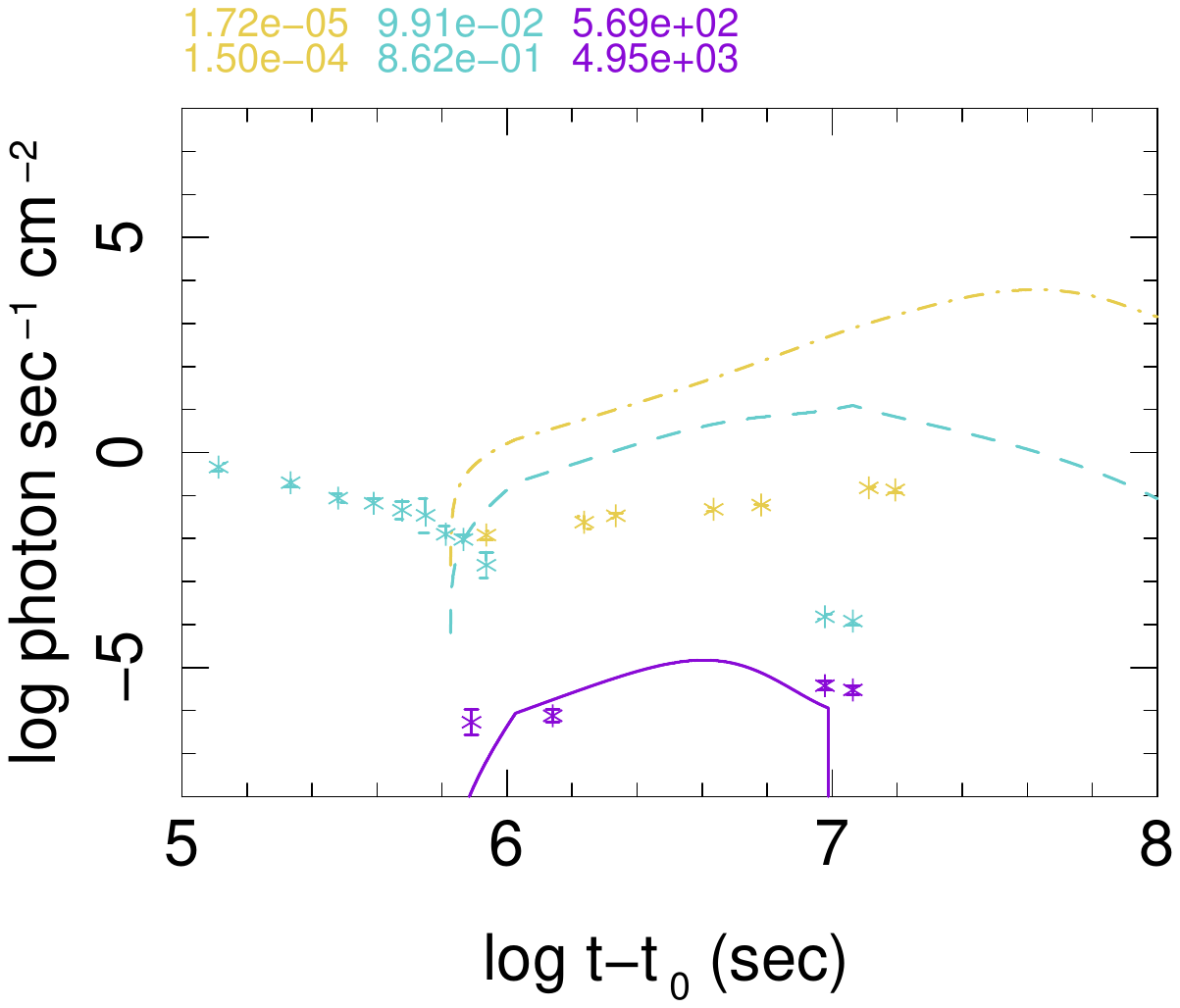} & 
\hspace{-1.5cm}\includegraphics[width=7cm]{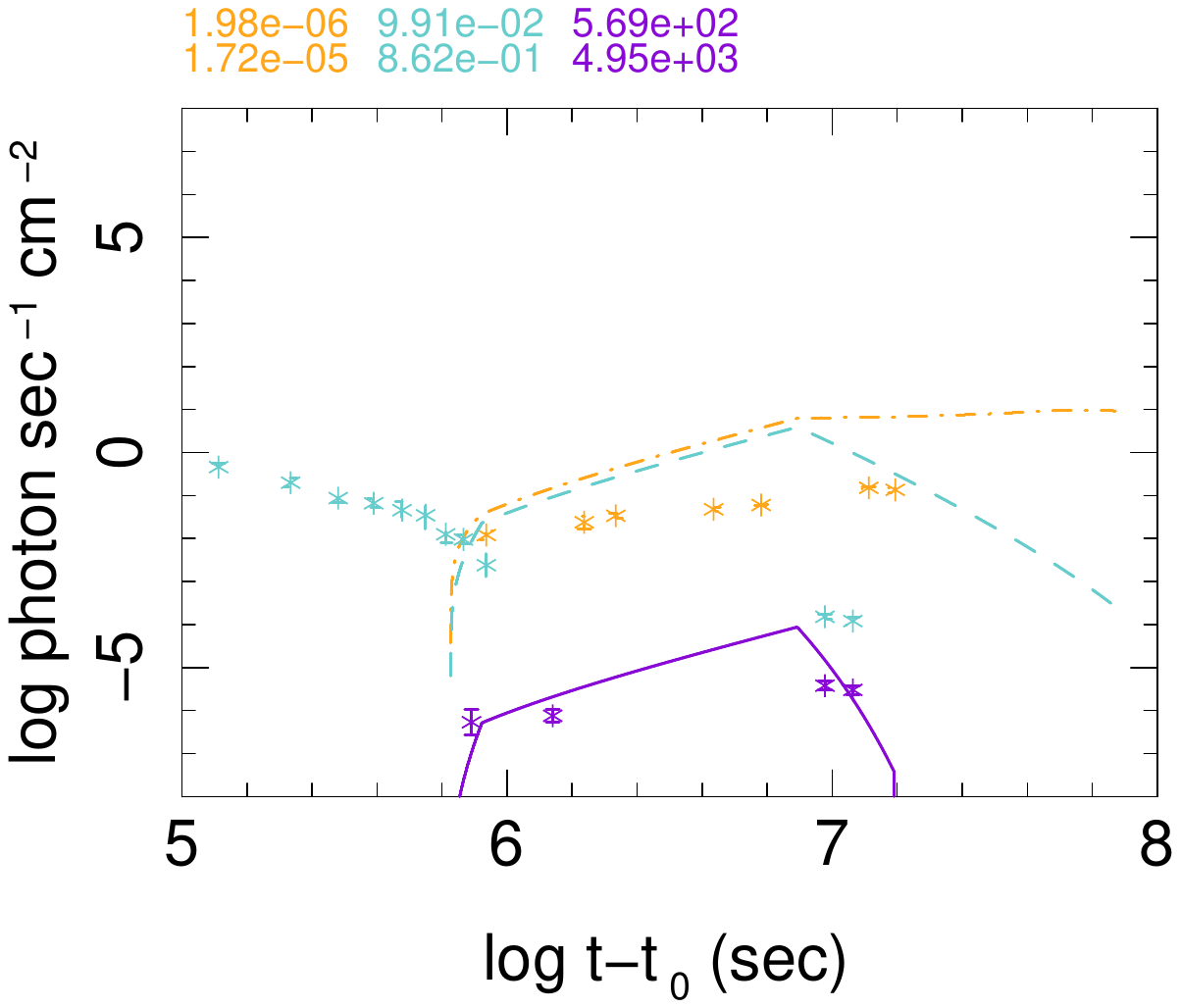} & 
\hspace{-3cm}\includegraphics[width=7cm]{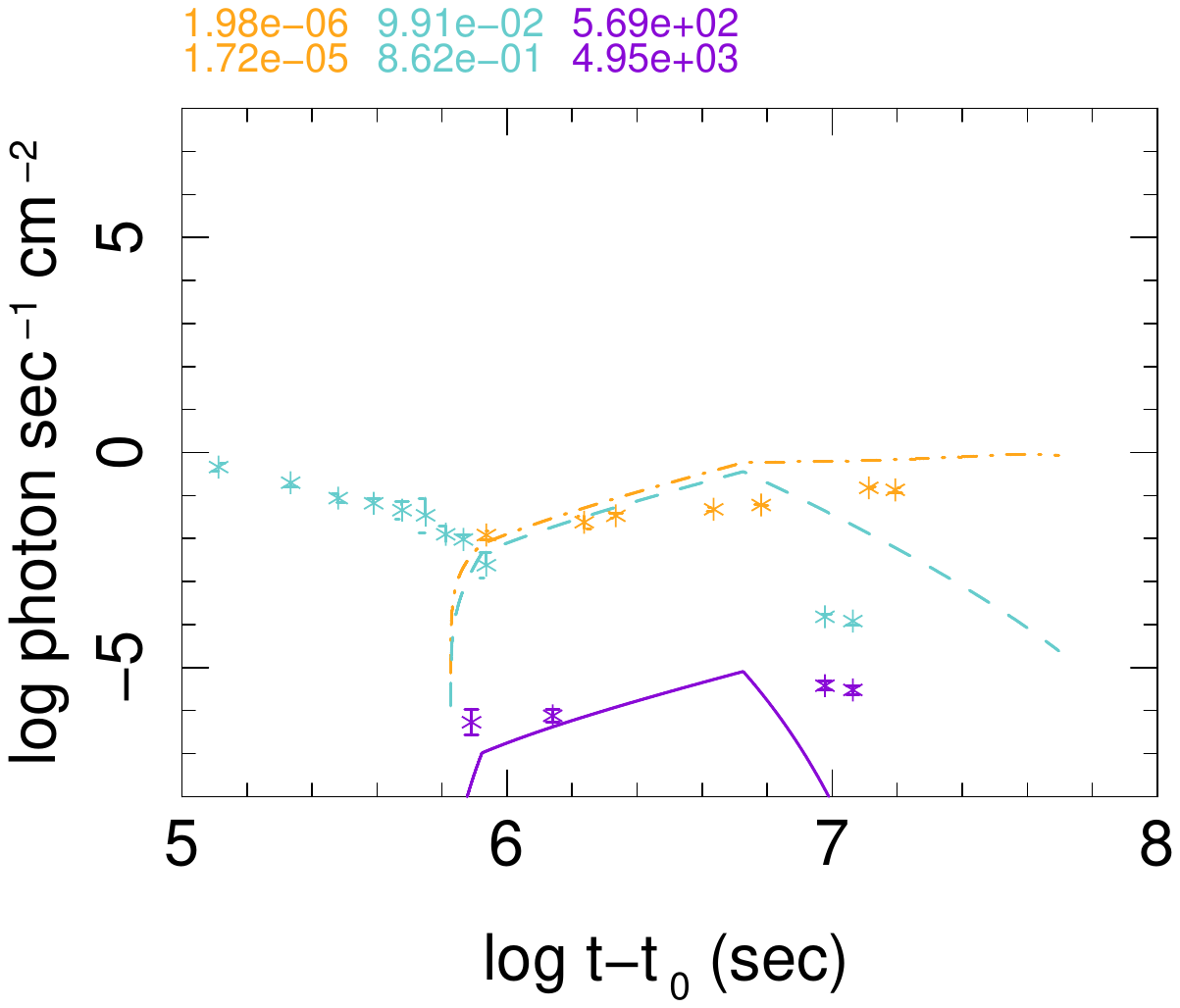} \\
 & & \\
Simul. 4 & \hspace{-1.5cm} Simul. 5 & \hspace{-3cm} Simul. 6 \\
\includegraphics[width=7cm]{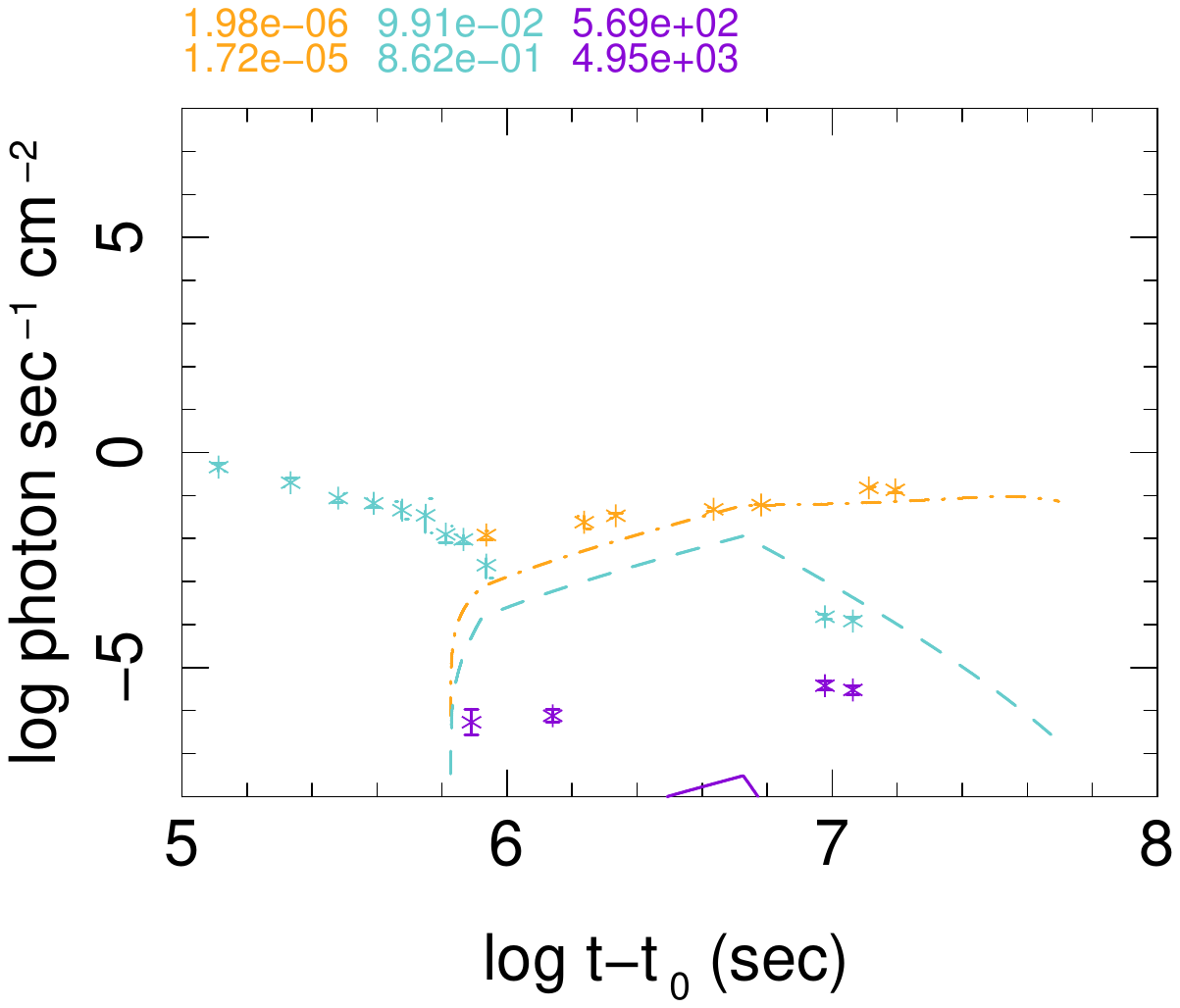} & 
\hspace{-1.5cm}\includegraphics[width=7cm]{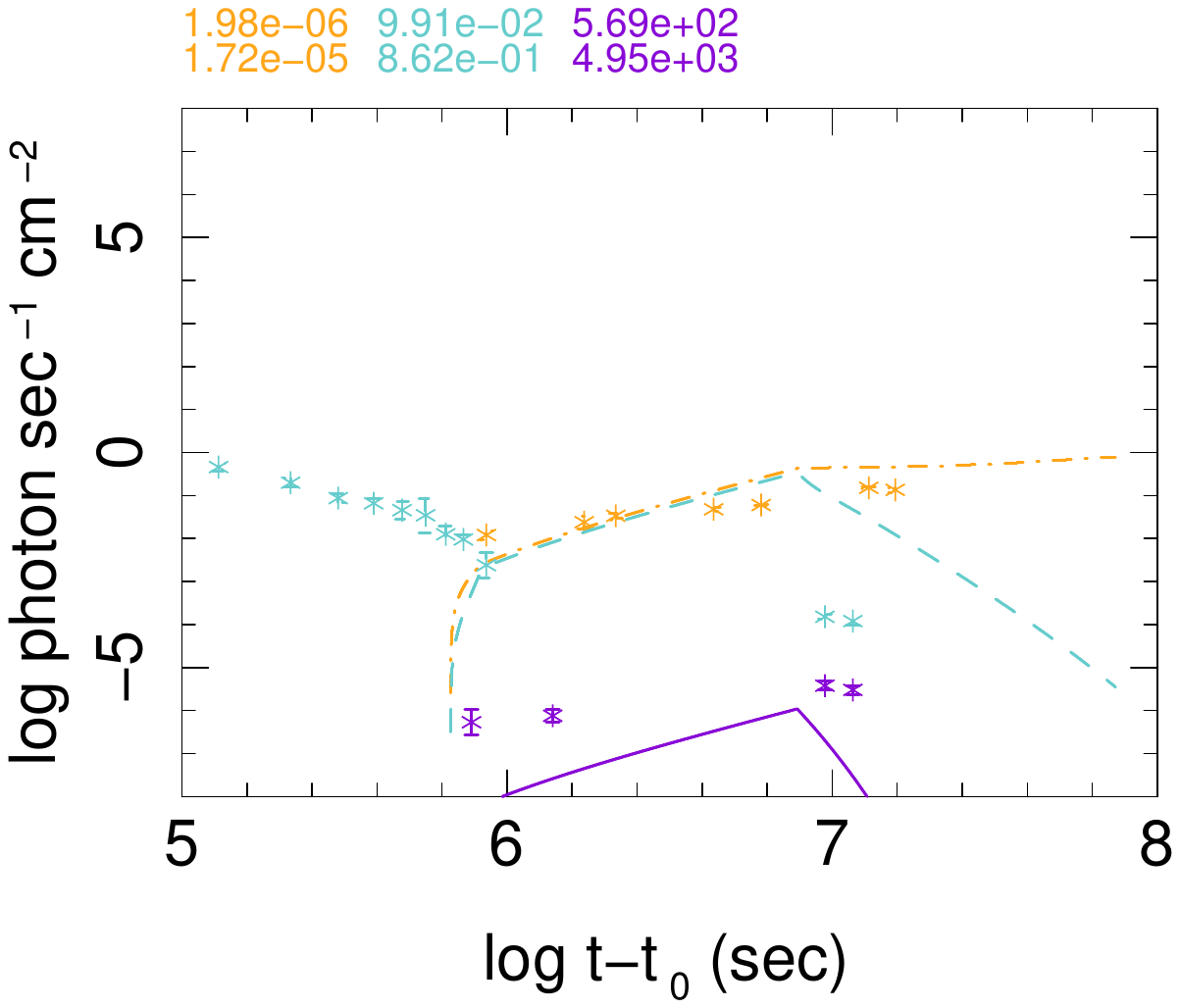} & 
\hspace{-3cm}\includegraphics[width=7cm]{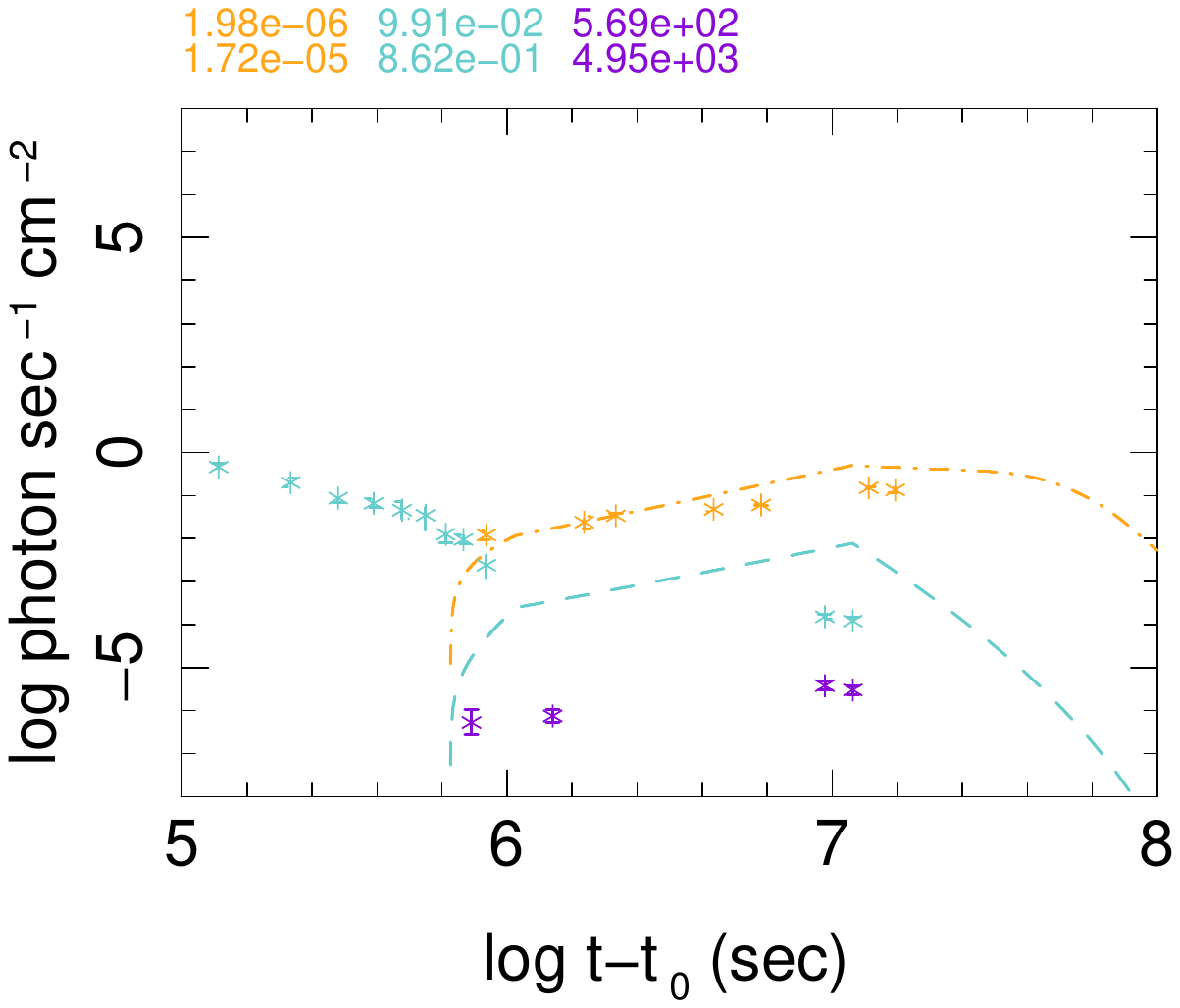} \\
 & & \\
Simul. 7 & \hspace{-1.5cm} Simul. 8 & \\
\includegraphics[width=7cm]{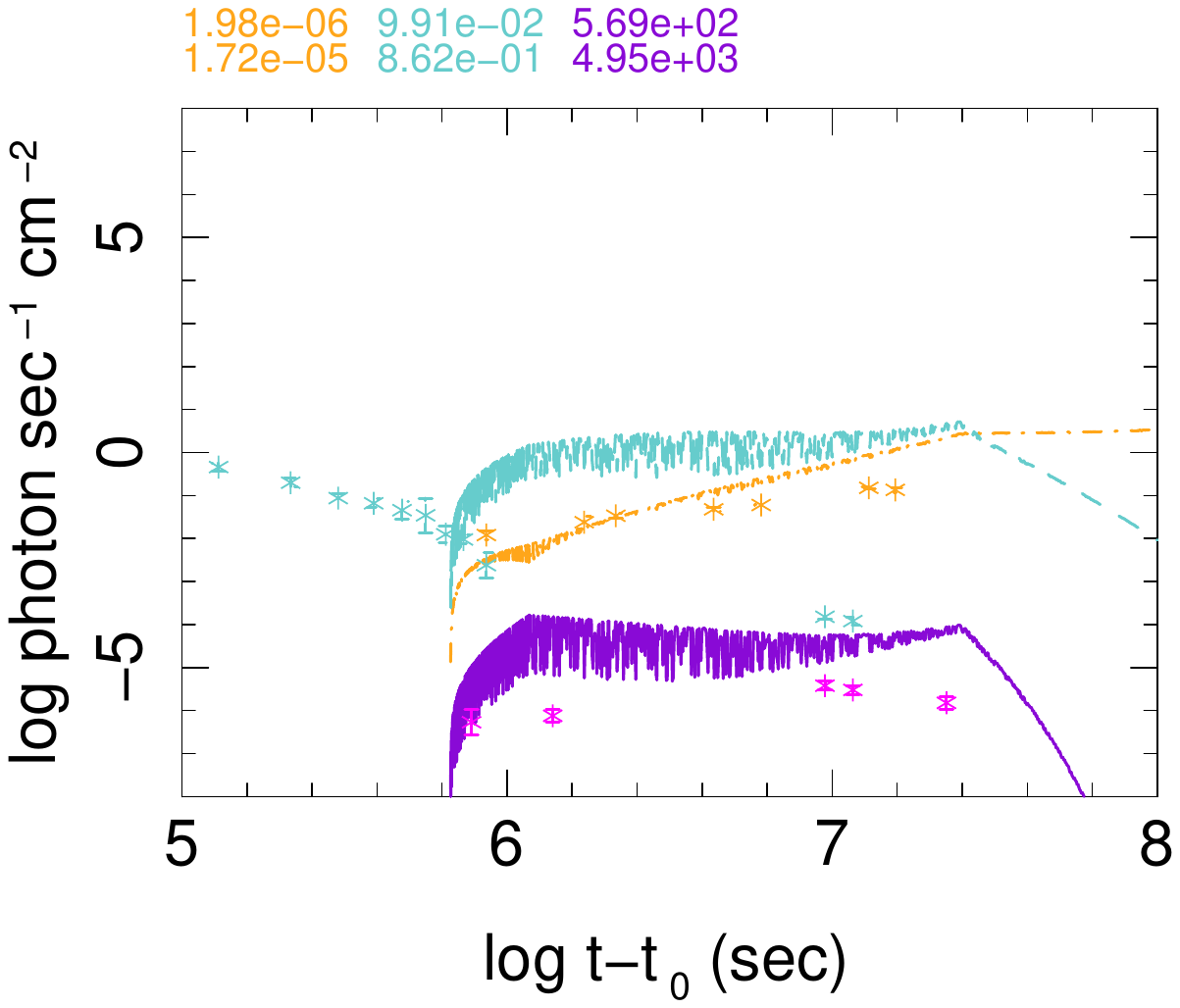} & 
\hspace{-1.5cm}\includegraphics[width=7cm]{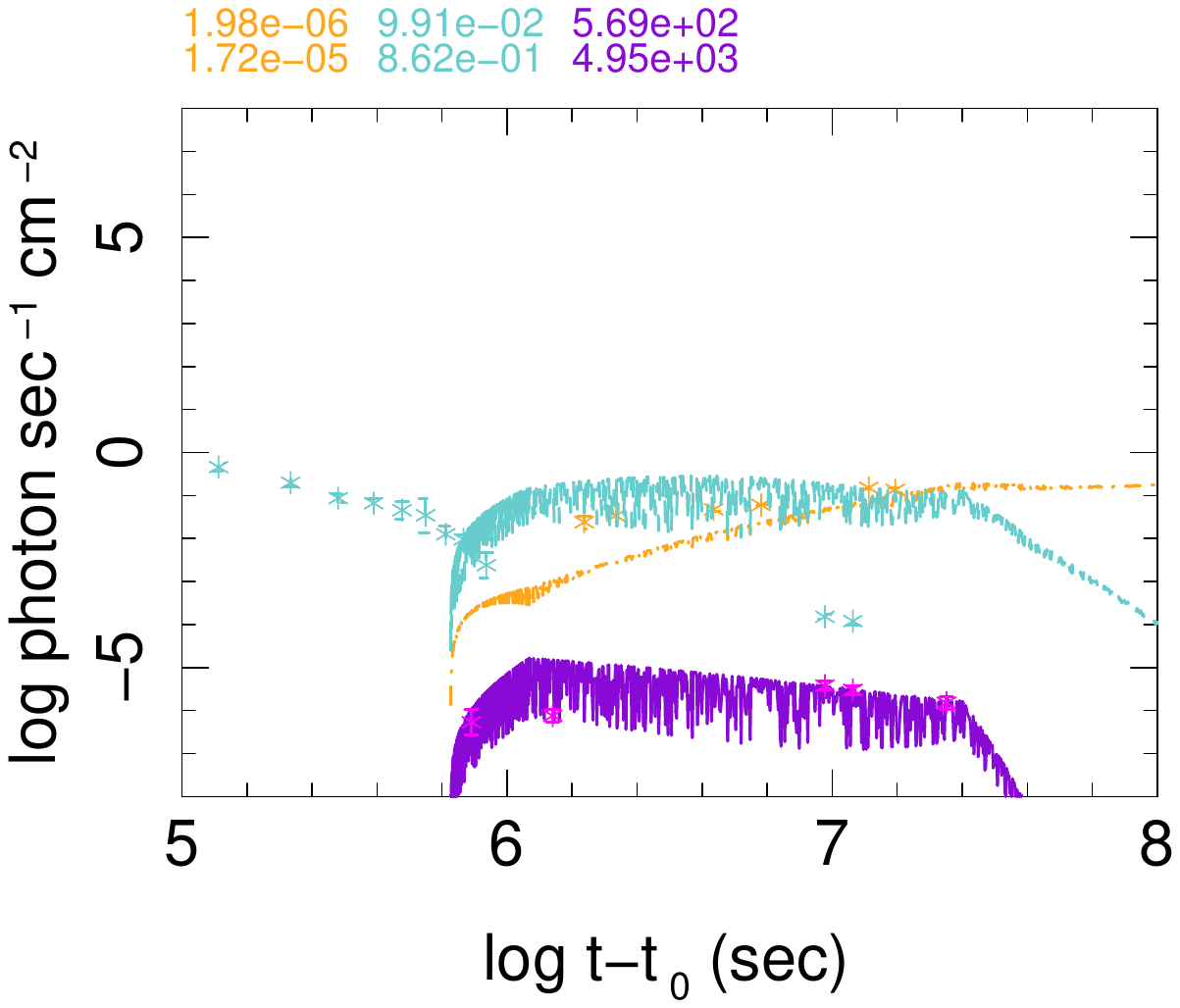} & 
\end{tabular}
\caption{X-ray, optical, and radio light curves of simulated models without taking into account 
synchrotron self-absorption. The energy range for each band is written in the same colour/gray scale 
as the curve on the top of each plot. Stars present data taken from:~\citep{gw170817latexary} 
(X-ray), \citep{gw170817optdes,gw170817rprocess,gw170817lateopt,gw170817latedecline} (optical), 
\citep{gw170817earlyradio1,gw170817earlyradio,lateradio} (radio). Parameters of simulations No. 1 to 6 corresponds to 
models No. 1 to 6 in Table \ref{tab:param}. Simulation No. 7 has the same parameters as simulation 
No. 4 but include an external magnetic field - presumably a Poynting flow oscillating with a 
frequency of $0.01$~Hz and $|B| = 5 (r/r_0)^{\alpha_m}$~G, where $\alpha_m = 1, 1, 2$ in the 3 regimes 
of this simulation. Simulation No. 8 corresponds to model No. 7 in Table \ref{tab:param} and 
includes an external magnetic field similar to that of simulation No. 7. Note that the plots of the 
last 2 simulations include X-ray observation by Chandra at 
$\sim T+260$~days~\citep{gw170817xraycxc260}. \label{fig:lc}}
\end{center}
\end{figure}

Fig. \ref{fig:lc} shows simulations which reproduce well at least one of the 3 energy bands with 
late observations, namely X-ray, optical/IR and 6~GHz radio bands. Parameters of these 
simulations are shown in Table \ref{tab:param}. We remind that only two relatively close in 
time data points are available in optical/IR band at late times\footnote{At the time 
of preparation of this article new observations of GW 170817 counterpart in far-IR at 
$\sim T + 264$~days was published~\cite{gw170817lateir}. They are not considered in our analysis, 
partly to avoid further delay in the completion of our work and partly because far-IR emission at 
the epoch of observations is expected to be dominated by kilonova emission~\cite{gw170817lateir}.}. 
Earlier data in this band is dominated by the kilonova emission and cannot be used to investigate 
the origin and properties of late emission, which as we discussed earlier, is presumably produced 
by other components of the ejecta from the BNS merger, in particular by a mildly relativistic outflow.
Values of Lorentz factor (or equivalently $\beta$) chosen for these simulations, namely 
$\beta = 0.4, 0.8$ correspond to what is used in the literature for simulating late light curves of 
GW/GRB 170817. As we discussed in Sec. \ref{sec:model}, we performed simulations with different 
Lorentz factor and flow density to mimic variation of outflow characteristics which couldn't be 
included in our simplistic model. 

%

\begin{table}
\begin{center}
\caption{Parameter set of simulated models. \label{tab:param}}
\end{center}
{\scriptsize
\begin{center}
\begin{tabular}{p{5mm}p{5mm}p{5mm}p{15mm}p{12mm}p{10mm}p{5mm}p{5mm}p{5mm}p{5mm}p{10mm}p{5mm}p{5mm}p{5mm}p{10mm}p{10mm}|}
\hline
Simul. No. & mod. & $\gamma'_0$ & $r_0$ (cm) & $\frac{\Delta r_0}{r_0}$ & $(\frac{r}{r_0})_{max}$ & $p$ & $\gamma_{cut}$ & $\kappa$ & $\delta$ & $\epsilon_B$ & $\alpha_B$ & $\epsilon_e Y_e$ & $\alpha_e$ & $N'$ (cm$^{-3}$) & $n'_c$ (cm$^{-2}$) \\
\hline 
\multirow{4}{5mm}{1} 
 & 1 & 2.3 & $2.34 \times 10^{16}$ & $10^{-5}$ & 1.5 & 1.8 & 100 & -0.1 & 0.5 & $5 \times 10^{-4}$ & -1 & 0.02 & -1 & $0.5 \times 10^3$ & $10^{22}$ \\
 & 2 & -   &   -      &  -    & 10  &  -  & 100 & -0.4 & 0.1 &  -     &  0 &  -   &  0 &   -    &    -   \\
 & 2 & -   &   -      &  -    & 50  &  -  & 100 &  0   & 0.5 &  -     &  1 &  -   &  1 &   -    &    -   \\
 & 2 & -   &   -      &  -    & 10  &  -  & 100 &  0.5 &  1  &  -     &  1 &  -   &  1 &   -    &    -   \\
\hline
\multirow{4}{5mm}{2} 
 & 1 & 1.2 &  $10^{16}$ & $5 \times 10^{-3}$ & 1.5 & 1.8 & 100 & -0.5 &  1  & 0.05   & -1 & 0.1  & -1 &   1 & $5 \times 10^{23}$ \\
 & 2 & -   &   -      &  -    & 15  &  -  & 100 &  0   & 0.1 &  -     &  0 & -    &  0 &   -    &    -   \\
 & 2 & -   &   -      &  -    & 10  &  -  & 100 &  1   &  1  &  -     &  1 & -    &  1 &   -    &    -   \\
\hline
\multirow{4}{5mm}{3} 
 & 1 & 1.2 &  $10^{16}$ & $10^{-3}$ & 1.5 & 1.8 & 100 & -0.5 &  1  & 0.05   & -1 & 0.1  & -1 &   1    & $5 \times 10^{23}$ \\
 & 2 & -   &   -      &  -    & 10  &  -  & 100 &  0   & 0.1 &  -     &  0 & -    &  0 &   -    &    -   \\
 & 2 & -   &   -      &  -    & 10  &  -  & 100 &  1   &  1  &  -     &  1 & -    &  1 &   -    &    -   \\
\hline
\multirow{4}{5mm}{4} 
 & 1 & 1.2 &  $10^{16}$ & $5 \times 10^{-3}$ & 1.5 & 2.1 & 100 & -0.5 &  1  & 0.08   & -1 & 0.1  & -1 &   0.08 & $5 \times 10^{23}$ \\
 & 2 & -   &   -      &  -    & 10  &  -  & 100 &  0   & 0.1 &  -     &  0 & -    &  0 &   -    &    -   \\
 & 2 & -   &   -      &  -    & 10  &  -  & 100 &  1   &  1  &  -     &  1 & -    &  1 &   -    &    -   \\
\hline
\multirow{4}{5mm}{5} 
 & 1 & 1.2 &  $5 \times 10^{16}$ & $10^{-3}$ & 1.5 & 2.5 & 100 & -0.5 &  1  & 0.05   & -1 & 0.1  & -1 &   0.5  & $5 \times 10^{23}$ \\
 & 2 & -   &   -      &  -    & 10  &  -  & 100 &  0   & 0.1 &  -     &  0 & -    &  0 &   -    &    -   \\
 & 2 & -   &   -      &  -    & 10  &  -  & 100 &  1   &  1  &  -     &  1 & -    &  1 &   -    &    -   \\
\hline
\multirow{4}{5mm}{6} 
 & 1 & 2.3 & $2.34 \times 10^{16}$ & $10^{-3}$ & 1.5 & 1.8 & 100 & -0.5 &  1  & 0.05  & -1 & 0.02 & -1 & 1.e-1  & $5 \times 10^{22}$ \\
 & 2 & -   &   -      &  -    & 10  &  -  & 100 &  0.2  & 0.1 &  -    &  0 & -    &  0 &   -    &    -   \\
 & 2 & -   &   -      &  -    & 10  &  -  & 100 &  1.5  &  1  &  -    &  1 & -    &  1 &   -    &    -   \\
\hline
\multirow{4}{5mm}{7} 
 & 1 & 1.2 &  $10^{16}$ & $5 \times 10^{-3}$ & 2.5 & 2.1 & 100 & -0.5 &  1  & 0.08   & -1 & 0.1  & -1 &   0.008 & $10^{22}$ \\
 & 2 & -   &   -      &  -    & 30  &  -  & 100 &  0   & 0.1 &  -     &  0 & -    &  0 &   -    &    -   \\
 & 2 & -   &   -      &  -    & 20  &  -  & 100 &  1   &  1  &  -     &  1 & -    &  1 &   -    &    -   \\
\hline
\end{tabular}
\end{center}
}
\begin{description}
\item{$\star$} For external shocks of jet with the ISM or circumburst material $\Gamma = 1$ is 
assumed. In this case $\gamma'$ is the Lorentz factor of jet with respect of a far observer at 
the same redshift.
\item {$\star$} Each data line corresponds to one simulated regime, during which quantities listed 
here remain constant or evolve dynamically according to fixed rules. A full simulation of a burst 
usually includes multiple regimes (at least two). 
\item {$\star$} Horizontal black lines separate time intervals (regimes) of independent simulations 
identified by the number shown in the first column.
\item {$\star$} A dash as value for a parameter presents one of the following cases: it is 
irrelevant for the model; it is evolved from its initial value according to an evolution equations 
described in~\citep{hourigrb,hourigrbmag}; or it is kept constant during all regimes. 
\end{description}
\end{table}

Simulations No. 1 and 2 are good fits to X-ray data, simulation 
No.3 fits X-ray and radio , and simulation No. 4 fits well optical and radio data. A main difference 
between simulations which fit X-ray data and those which fit radio data is the higher density of 
ISM/circum-merger material in the former with respect to the latter. They presumably present a 
denser front part, which collides with a denser circum-merger material, and a diluted hind part, 
which collides with the remnant of ISM/circum-merger material after its distortion and spread out 
by the front of the flow, respectively. Apriori a linear interpolation between these simulations 
should lead to an acceptable fit of all three energy bands. However, it is easy to see that no 
linear combination of simulations fits all the light curves and either leads to over production of 
optical and radio emissions by the denser front part of the flow or under-estimation of X-ray. 
Indeed an interpolation in time between simulations with high and low Lorentz factors needs zero 
contribution from the denser part of outflow around the time of last X-ray observations to be 
consistent with optical observations at the same time.

We also considered the effect of a Poynting flow imprinted in the outflow material. Ejected material 
from BNS merger is expected to be initially highly magnetized and may preserve part of its magnetic 
energy well after its expansion. Even in the case of main sequence stars such as the Sun a weak 
magnetic field is detected at upstream of wind termination shock~\citep{solarwind}. Simulations 
No. 7 and 8 in Fig. \ref{fig:lc} are examples of the case of a magnetically loaded outflow. Simulation 
No. 7 has the same parameters as Simulation No. 4, which fits radio and optical data well but has 
insufficient X-ray. We find that when a magnetic field with initial flux of $|B| \sim 5$~G is added 
to this model, both X-ray and optical emissions become much brighter than observations. Simulation 
No. 8 in Fig. \ref{fig:lc} has the same parameters as model No. 7 in Table \ref{tab:param}. It 
presents a more diluted outflow and circum-burst material but the same magnetic field as simulation 
No. 7. Fig. \ref{fig:lc} shows that its X-ray light curve is consistent with the data. Moreover, 
the lower column density in this model is better consistent with the estimated value found in 
Sec \ref{sec:cocoon}. However, its optical emission is too bright and similar to simulations No. 1 
to 3. Thus, only in presence of significant extinction the model this simulation can be consistent 
with data. In both these examples radio emission remains consistent with observations.

To explain discrepancies between data and simulated optical and radio light curves we first consider 
two important processes which are so far overlooked. They are synchrotron self-absorption in radio 
band and extinction of optical emission by material around the merger and in its environment. 


\subsection{Absorption} \label{sec:abs}
Low energy emissions are absorbed by gas and dust in circumburst material and ISM, and through 
synchrotron self-absorption process in the shocked region of the outflow. The main source of 
absorption for X-ray is neutral atomic gas. The $H_I$ equivalent column density of the Milky Way 
$N_H^{MW}$ in the direction of GW/GRB 170817A is 
$7.84 \times 10^{20}$~cm$^{-2}$~\citep{gw170817cxc2day,gw170817xray}. Using 0.3-10~keV X-ray data from 
Chandra, the intrinsic column density in front of the ejecta of GW/GRB 170817A is estimated to be 
$\lesssim 3 \times 10^{22}$~cm$^{-2} \approx 0.05$~gr/cm$^2$~\citep{gw170817cxc2day,gw170817xray}. It 
presumably presents material in the local environment of the progenitor BNS and is much smaller than 
characteristic absorption length of 0.3-10~keV X-ray, which is 
$\sim 0.2$~gr/cm$^2$~\citep{particledata}. Thus, X-ray absorption is only a few 
percents~\citep{particledata} and negligible with respect to observational and modelling 
uncertainties.

\begin{figure}
\begin{center}
\begin{tabular}{p{5.5cm}p{5.5cm}p{5.5cm}p{5.5cm}}
a) & \hspace{-1cm} b) & \hspace{-2cm} c) & \hspace{-3cm} d) \\
\includegraphics[width=5.5cm]{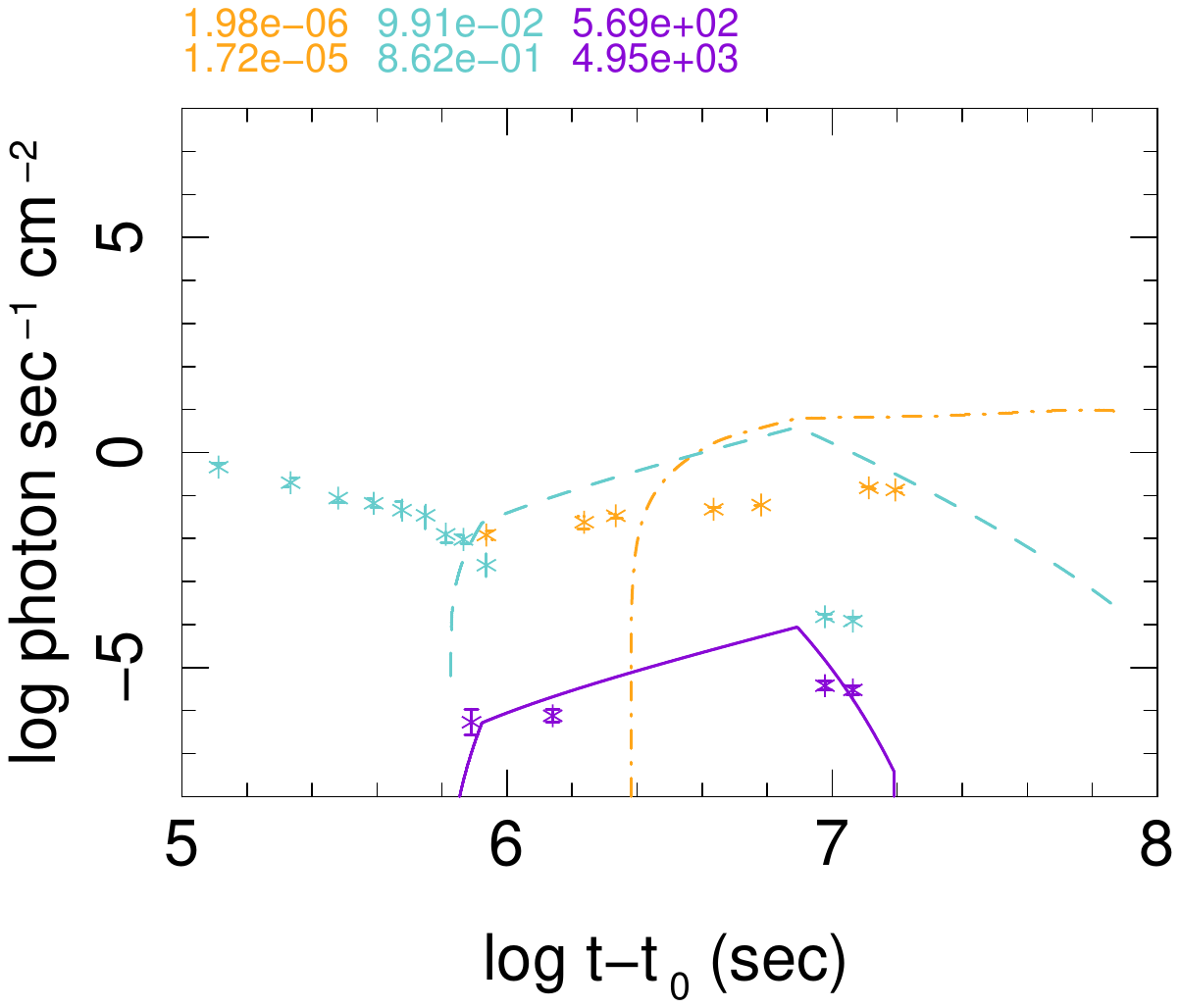} & 
\hspace{-1.5cm}\includegraphics[width=5.5cm]{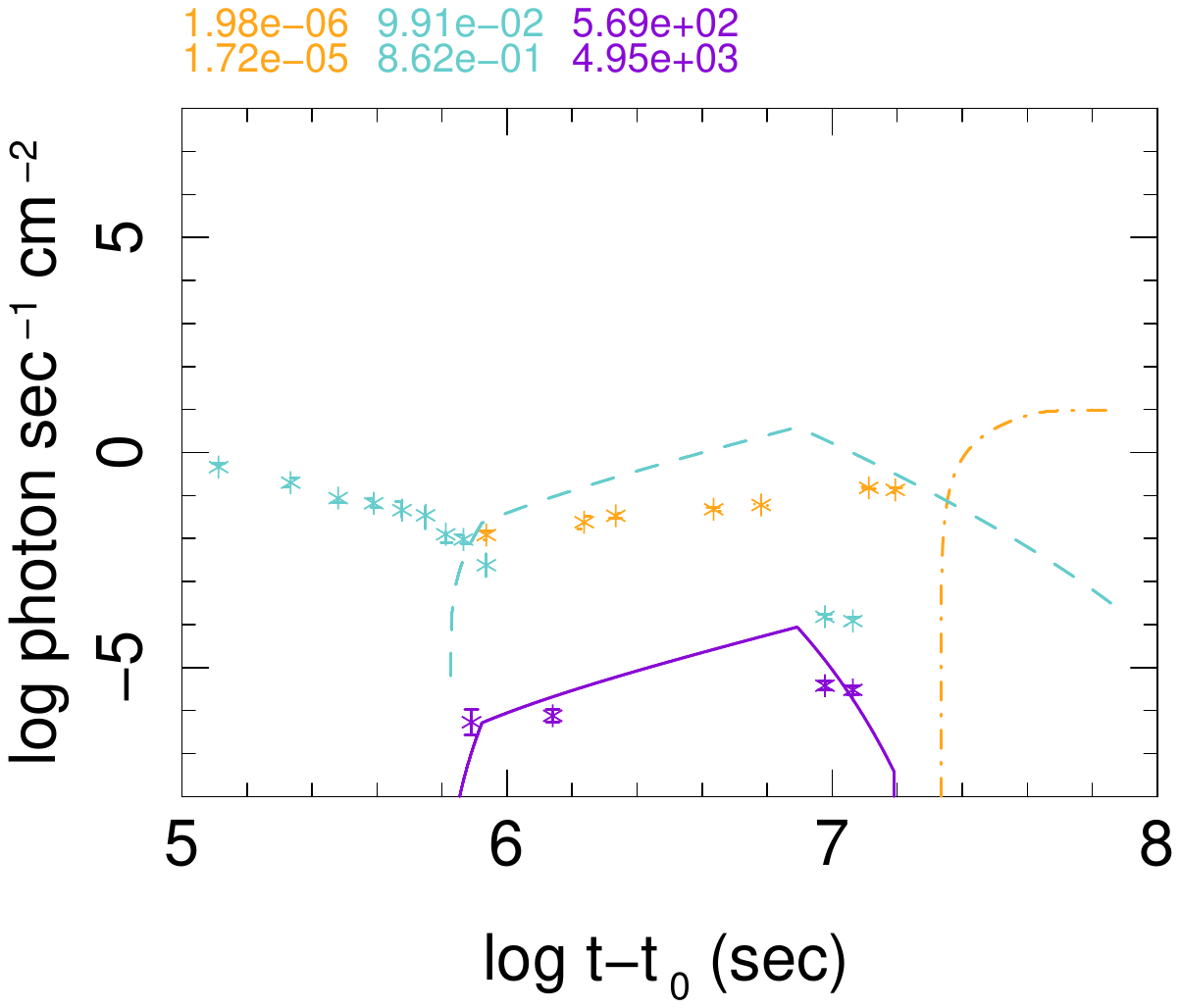} & 
\hspace{-3cm}\includegraphics[width=5.5cm]{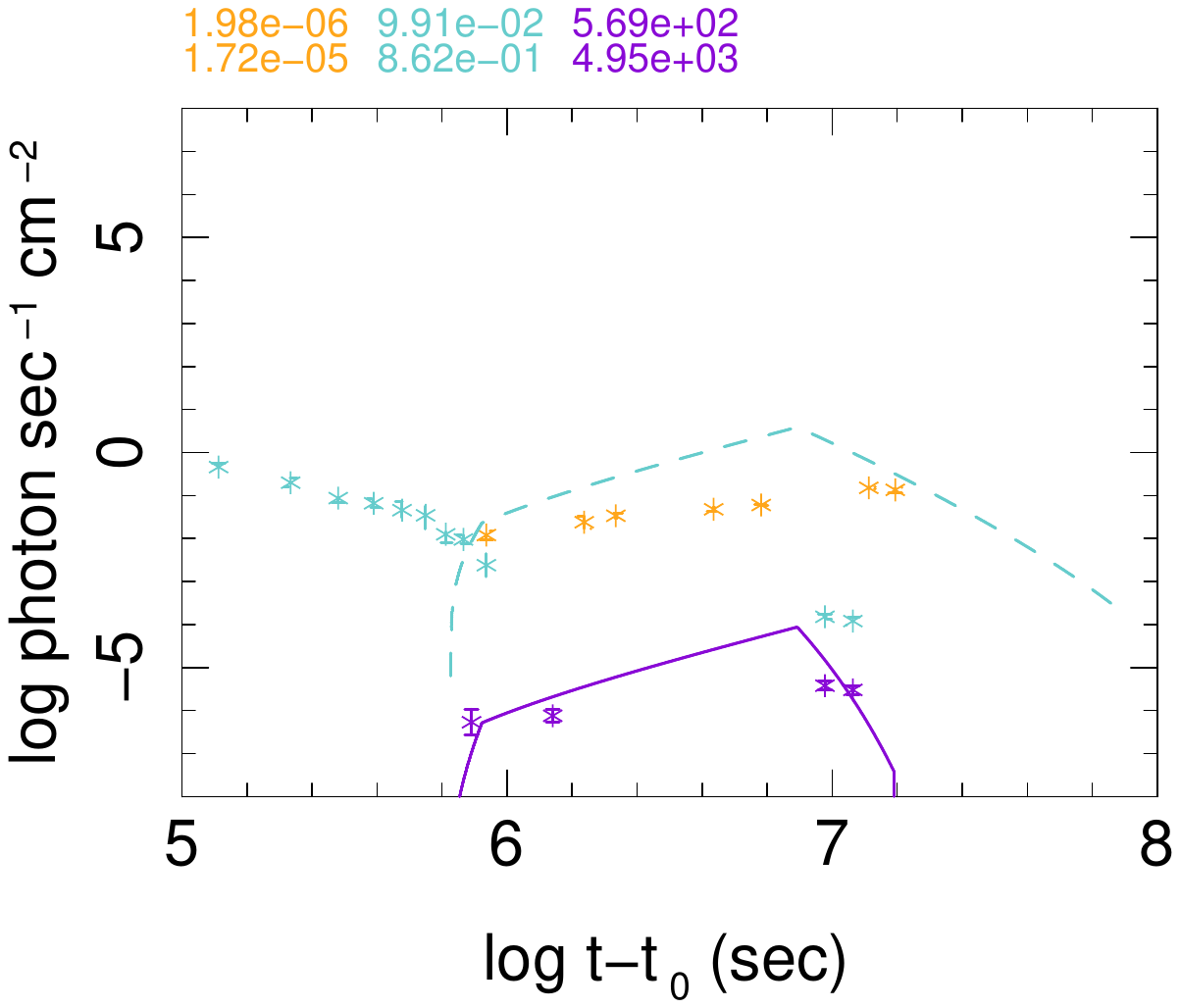} &
\hspace{-4.5cm}\includegraphics[width=5.5cm]{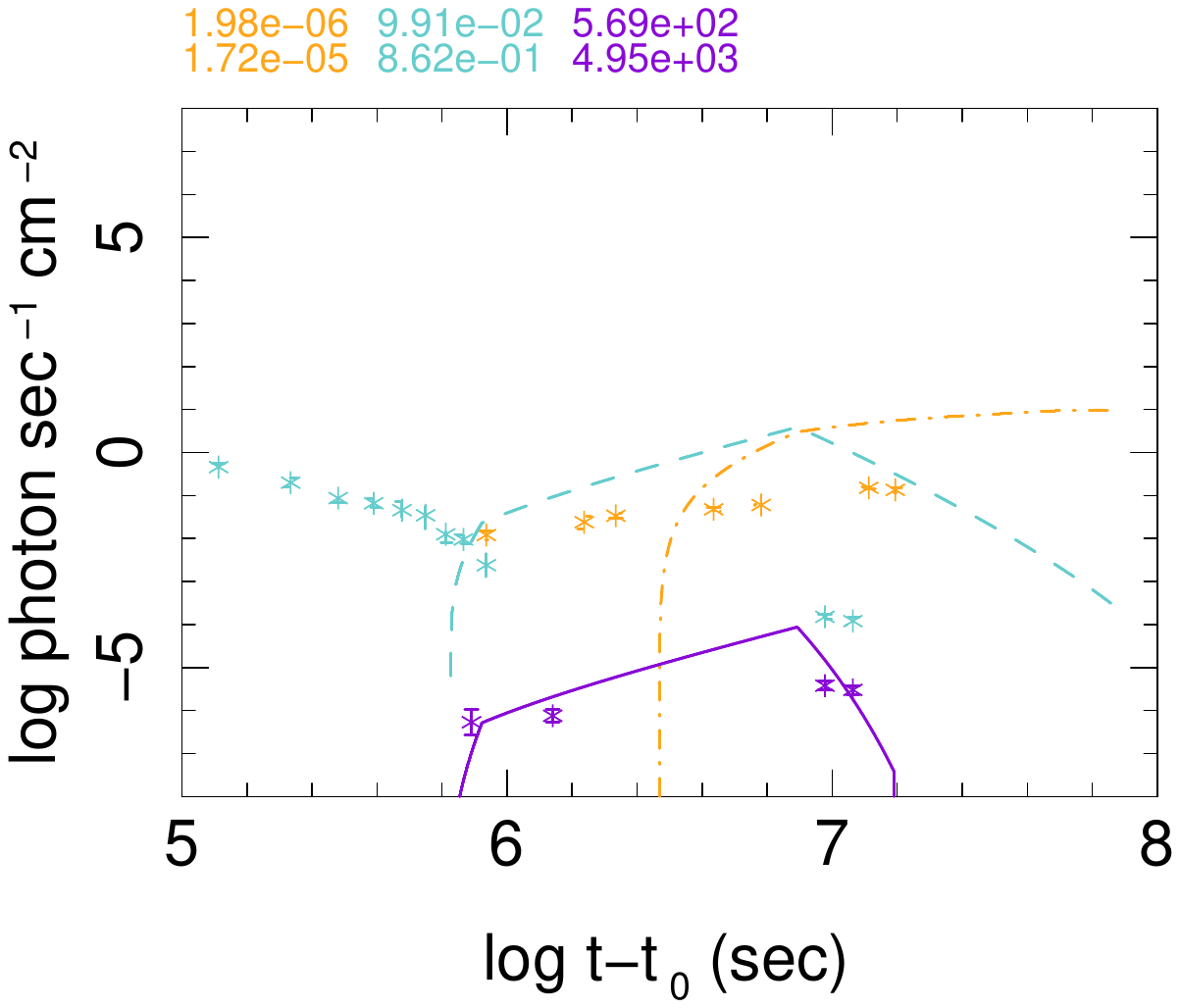} 
\end{tabular}
\caption{Light curves of Model No. 2 with different self-absorption length index $\alpha$ defined 
in Appendix \ref{app:abs}: Exponential dependence on distance and $\alpha = 1$ (a), $\alpha = 0.1$ 
(b), and $\alpha = 0$ (c); power-law dependence on distance and $\alpha = 3$ (d). Description of 
light curves and data is the same as Fig. \ref{fig:lc}\label{fig:lcabs}}
\end{center}
\end{figure}

In astronomical shocks synchrotron self-absorption is significant only in radio 
band~\citep{emissionbook}. In appendix \ref{app:abs} we calculate synchrotron self-absorption 
coefficient in the framework of phenomenological shock model of~\citep{hourigrb}. Plots in 
Fig. \ref{fig:lcabs} show the effect of synchrotron self-absorption for simulation No. 2 and a 
phenomenological description for the variation of absorption length with propagation as an 
exponential or power-law, see Appendix \ref {app:abs} for details. They confirm that 
$\lesssim 6$~GHz radio emission can be completely absorbed if the extension of the induced magnetic 
field in front of the shocked region is enough long. Indeed, radio emission in both short and 
long GRBs is only observed at late times and well after the decline of higher energy counterparts. 
In the case of GW/GRB 170817A the first detection of a counterpart in radio was only a few days 
after the first detection in X-ray~\citep{gw170817earlyradio,gw170817earlyradio0,gw170817earlyradio1,gw170817lateradio,lateradio,gw170817latexoptradio1}
\footnote{Observations in this work was made public after our simulations and are not taken into 
account in our analysis. Nonetheless, they are consistent with our simulations and their 
interpretations.}. However, as it is explained in details in~\citep{hourigw170817} we have most 
probably missed the early X-ray and optical afterglows. Assuming that the denser front of the 
outflow had a long effective length for synchrotron self-absorption, its radio emission was 
completely extinguished locally and what we observe comes from the weak shocks of slower (tail) 
component of the flow presented by simulations No. 4, 5, and 6 in Fig. \ref{fig:lc}.

Optical and IR bands are mainly affected by dust and electrons in the outer shells of neutral or 
slightly ionized atomic gas. The extinction in the Milky Way in the direction of GW 170817 is 
estimated to be only $A_V \lesssim 0.5$ mag~\citep{gw170817optsss1}. Moreover, due to the very low 
star formation rate of the host galaxy NGC 4993, the extinction inside the host is believed to be 
negligible~\citep{gw170817bluekilonovapol}. Therefore, any further extinction must be local and due 
to the circum-merger material and environment. As mentioned earlier, X-ray data shows the existence 
of an equivalent $N_H \lesssim 3 \times 10^{22}$~cm$^{-2}$ column of material in front of the outflow. 
If this material were genuinely hydrogen, it had negligible effect on the 
optical/IR emission. However, in the environment of old collapsed stars a large fraction of this 
material should consist of heavier elements such as $O$, $C$, $Fe$, $Si$, etc.~\citep{nhavcal}. They  
can significantly affect low energy photons if they are not fully ionized. Indeed, measurements of 
equivalent column density $N_H$ and extinction $A_V$ in supernovae remnants show a linear relation 
between $\log N_H$ and $A_V$, see e.g.~\citep{nhextinctrel,nhextinctrel0,nhextinctrel1}. A recent 
calibration of this relation in SNRs~\citep{nhavcal} estimates this relation as 
$N_H = (2.1 \pm 0.09) \times 10^{21} A_V~(\text{cm}^{-2})$. Using this equation, we find that for 
$N_H \lesssim 3 \times 10^{22}$~cm$^{-2}$ the amount of extinction in visible band is 
$A_V \lesssim 10$~mag $\sim 4$~dex. This amount of extinction can make optical light curve of 
simulations No. 2, 3, and 8 fully consistent with the data.

\section{Discussion} \label{sec:discuss}
\begin{figure}
\begin{center}
\begin{tabular}{p{6cm}p{8.5cm}p{6cm}}
a) & b) & \hspace{-2.5cm} c)\\
\hspace{-1cm} \includegraphics[width=6cm,angle=90]{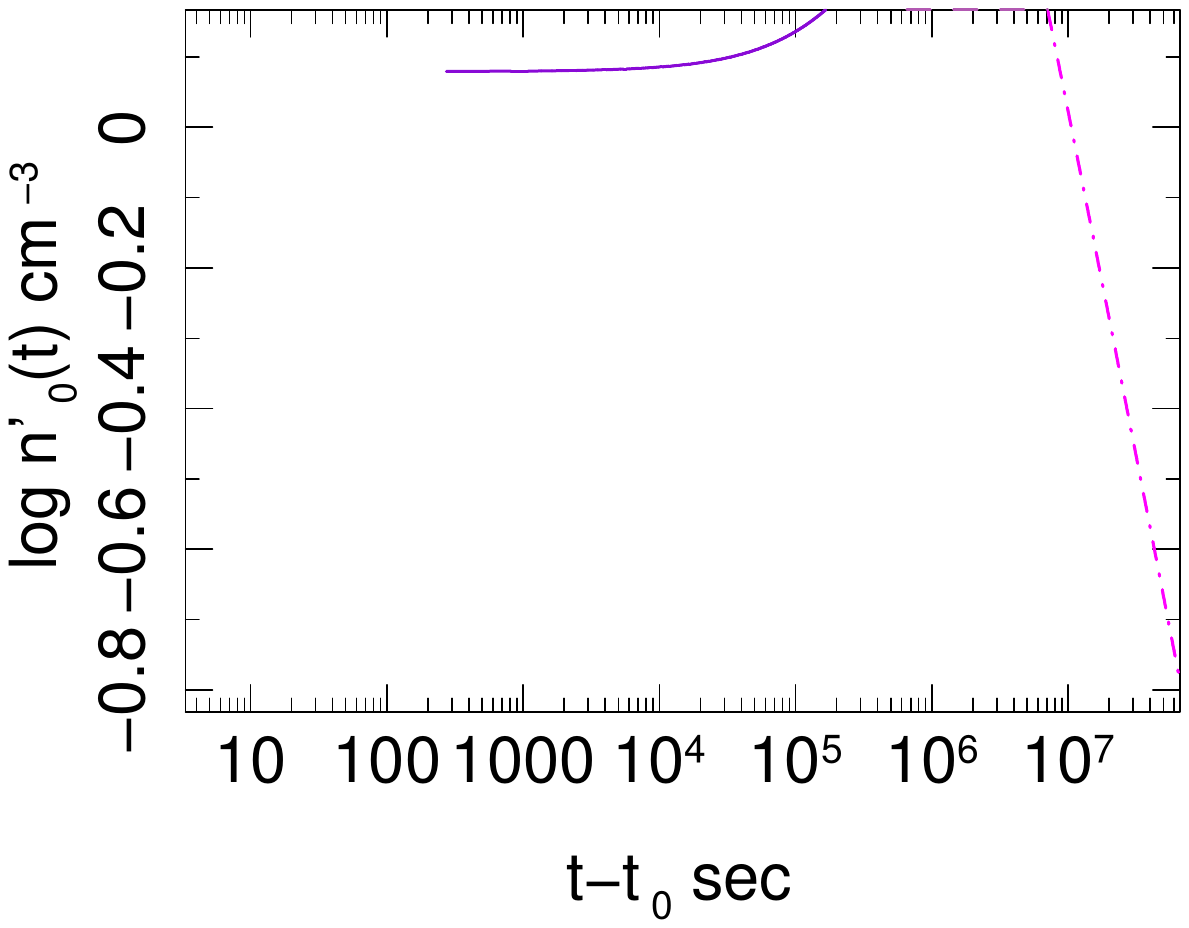} & 
\hspace{-1cm}\includegraphics[width=8.5cm]{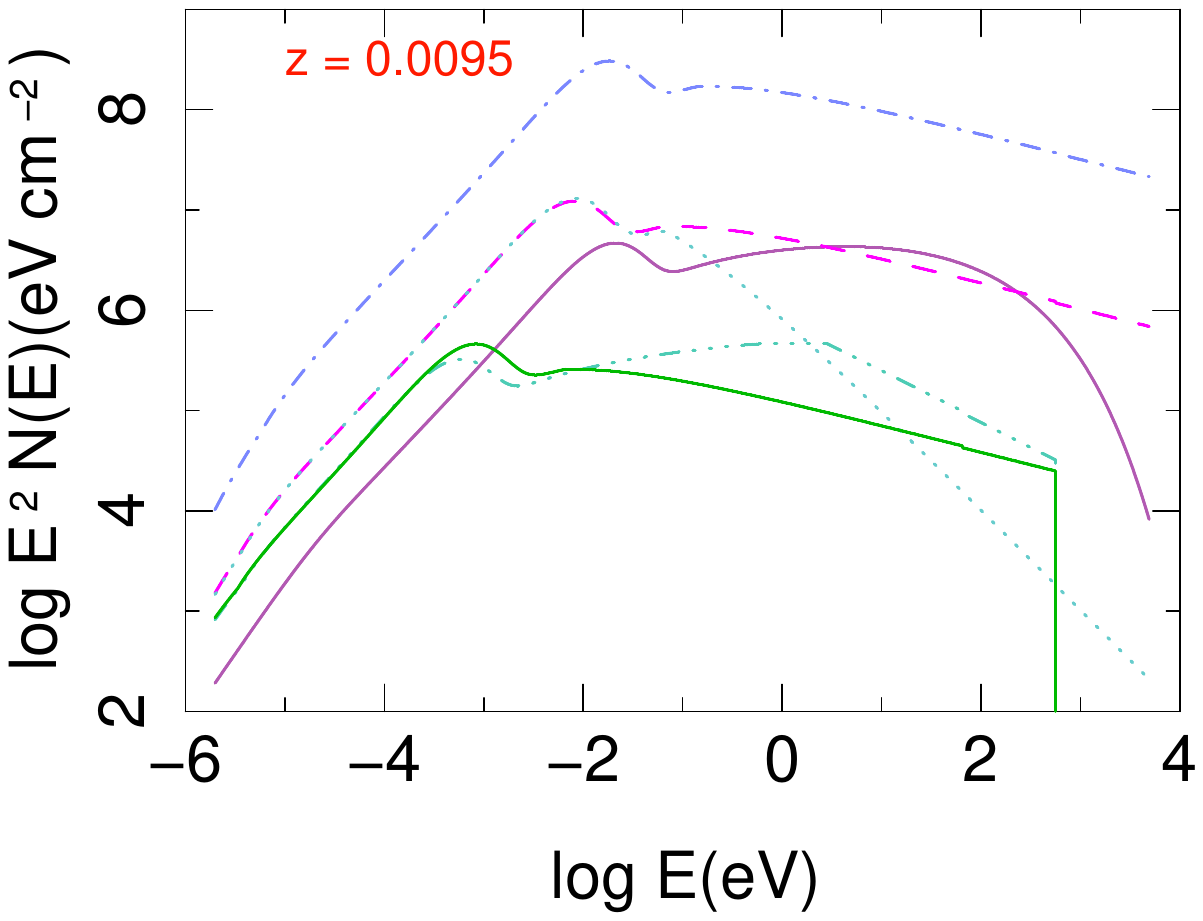} & 
\hspace{-3.5cm} \includegraphics[width=6cm,angle=90]{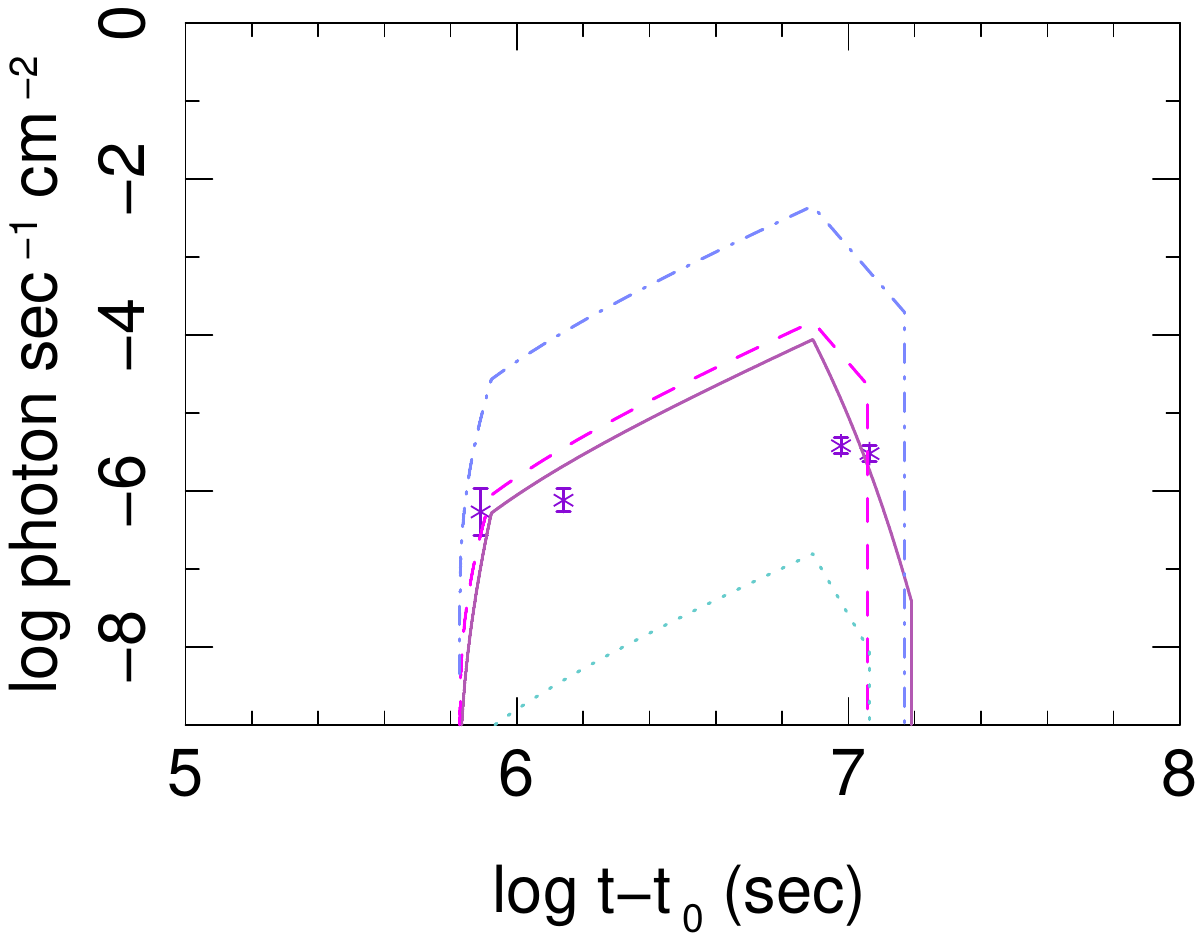} 
\end{tabular}
\caption{a) Profile of the material sheath around the BNS merger. b) Spectra of simulation No. 2 
(dark continuous) and several of its variants with broken power-law spectrum for electrons: 
$p_2 = 2.5$, $n'_0 = 1$~cm$^{-3}$ and $\Gamma = 3$ (dashed); the same as the previous with 
$n'_0 = 5$~cm$^{-3}$ (dash-dot); $p_2 = 2.5$ and $n'_c = 5 \times 10^{22}$~cm$^{-2}$ (dash-dot); 
$p_2 = 3$ (dash-3 dot); and $p_2 = 4$, $n'_0 = 1$~cm$^{-3}$ and $\Gamma = 3$ (light continuous). 
 c) X-ray light curves of models in b). The last two models in b) do not have significant emission 
in this band due to a noticeable break in their spectra at $\sim 0.5$~keV because the simulation 
code was not able to follow the evolution of rapidly fainting X-ray. \label{fig:n0}}
\end{center}
\end{figure}

To generate a light curve consistent with observations we assumed a density profile for the 
ISM/circumburst material. An example of such profiles is shown in Fig. \ref {fig:n0}-a. It evolves 
with the distance from central object according to a power-law with index $\kappa$ defined in 
Table \ref{tab:paramdef}. Its value for each model can be found in Table \ref{tab:param}. This 
phenomenological model allows to determine the profile of circumburst matter dynamically. It was 
presumably formed during the lifetime of the progenitor neutron stars~\citep{nssheath} and its 
compression in front of the outflow was responsible for generating a density discontinuity and a 
shock. One may argue that dynamical kick during supernova explosion of the progenitor massive stars 
might have kicked and ejected the progenitor neutron stars of GW/GRB 170817A event out of their 
birth place. Although this is a possibility, dynamical kick is not an inevitable outcome of massive 
stars collapse. Supernova remnants surrounding many neutron stars and pulsars in the Milky Way is 
an evidence for this fact~\citep{pulsarsnr}. Moreover, as the probability of the formation of BNS 
is higher in star clusters, even a moderate kick might not be sufficient to eject the neutron star 
out of the potential well of the cluster, which may be dustier than field~\citep{starclustdust}. 

Simulations shown in Fig. \ref{fig:lc} demonstrate that the density of matter sheath surrounding 
the BNS at the initial position of the shock front in the simulations must be at least few fold 
larger than average ISM density of the host galaxy reported in the literature, 
namely $\sim 0.04$~cm$^{-2}$ - concluded from the absence of significant neutral hydrogen in the 
host~\citep{gw170817earlyradio}. For lower densities the X-ray flux would be too small. 

Although due to simplifications described in Sec. \ref{sec:model} theoretical uncertainties of 
simulated models may be significant, general aspects of their spectra seem correct, 
see Fig. \ref {fig:n0}-b for some examples (and corresponding X-ray light curves in 
Fig. \ref {fig:n0}-c). Therefore, it is unlikely that the difficulty of simulations to explain 
multi-band observations without significant absorption of low energy photons and/or excess of X-ray 
be simply due to theoretical shortcomings of the model. Indeed, Fig. \ref {fig:n0}-b shows that for 
$\Gamma \lesssim 3$, ISM/circumburst density of $\mathcal{O} (1)$, and column density of flow as 
estimated in Sec. \ref {sec:cocoon}, the optical/IR band falls in the high energy wing of the 
spectrum. Thus, the only way to reduce the amount of energy emitted in optical band is to increase 
the peak energy. This cannot be achieved without much higher Lorentz factor and densities, which 
are in contradiction with what is expected from a mildly relativistic 
outflow~\cite{gw170817cocoonsimul} and a low density ISM/circumburst material, as observations of 
the host galaxy shows~\citep{gw170817earlyradio}. 

\begin{figure}
\begin{center}
\includegraphics[width=8cm]{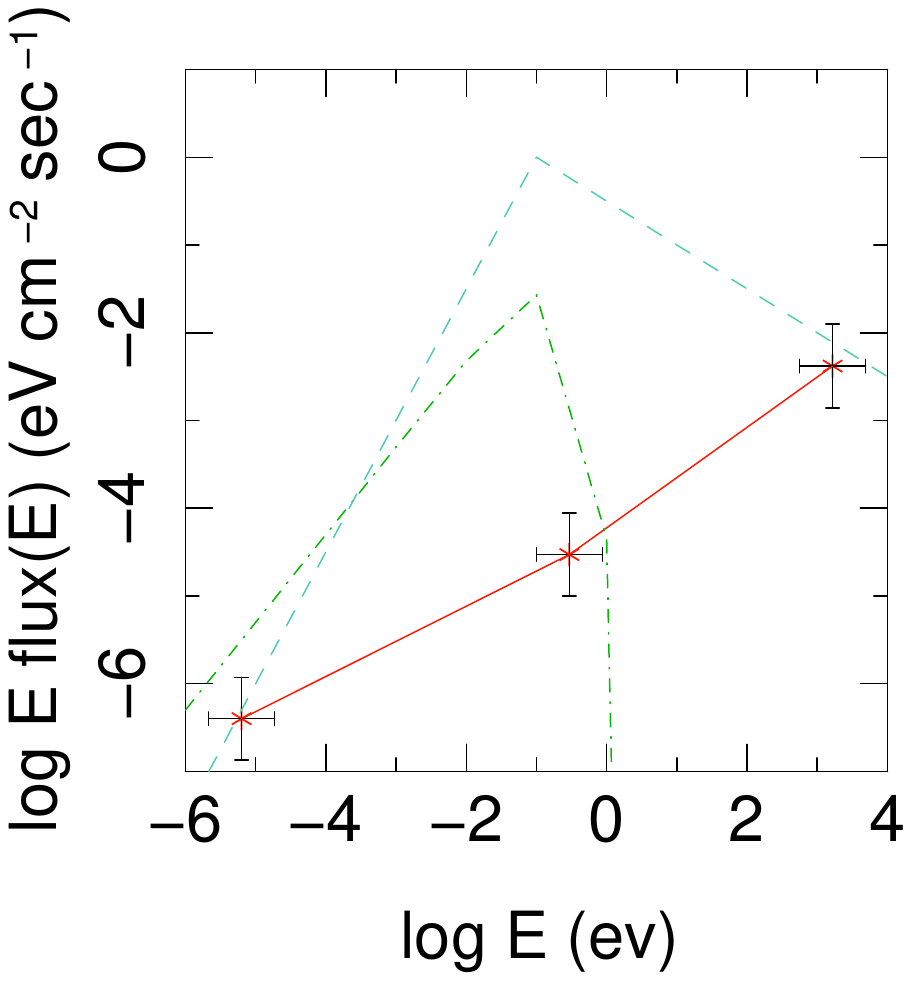}
\caption{Energy flux distribution of GW/GRB 170817A at $\sim T+110$~days. Bars on the data points 
present the width of corresponding energy band rather than measurement errors, which are much 
smaller. A typical synchrotron spectrum (dashed) is sketched by its asymptotic behaviour with 
theoretical slope of 3/2 at low energy wing and a chosen slope of -1/2 for high energy 
wing. Its peak is adjusted to a value close to simulated spectra shown in Fig. \ref{fig:n0}-b. 
A thermal spectrum with the same temperature as peak energy of the synchrotron spectrum is also 
shown (dash-dot). The continuous line connecting data points is drawn as guide for comparison with 
other curves. \label{fig:dataspect}}
\end{center}
\end{figure}

Irrespective of the results of our simulations, interpretation of multi-band observations as 
synchrotron or thermal emission would be very difficult without a significant optical/IR extinction 
or an excess of X-ray emission. Fig. \ref{fig:dataspect} shows the distribution of energy flux 
density at $\sim T + 110$~days, for which data in all three energy bands discussed here is 
available. We remind that this distribution is independent of viewing angle and geometry of 
emitting surface, which are not well understood and are subject to debate and controversies. 
It depends only on the emission mechanism. The observed distribution significantly deviates from a 
typical synchrotron and thermal spectra shown in Fig. \ref{fig:dataspect} or even a more realistic 
nonlinear + synchrotron + thermal spectrum simulated by~\citep{grbspectfermi,grbagspect}. It is not 
possible either to assume that all the three energy bands fall on the low energy wing of a 
synchrotron spectrum, because positive slope of the observed spectrum is much flatter than 3/2 of 
synchrotron emission. Moreover, if this assumption were true, the peak of spectrum had to be at 
even higher energies and the source should have been observable in hard X-ray and gamma-ray.

A large extinction of optical photons is consistent with the fact that optical afterglow of the 
majority of short bursts are not detected, most probably due to the intrinsic 
absorption~\citep{grbshortenv,uvotcat}. Assuming that host galaxies of short GRBs have gas column 
densities comparable to those of Milky Way and NGC4993, their contribution to equivalent $N_H$ should 
be at most few percent and the rest must be the contribution of immediate environment of source or 
the outflow itself. On the other hand, the upper limit on the $N_H$ obtained from X-ray observations 
and its associated extinction are consistent with those of globular clusters, 
see e.g.~\citep{globclustdens}. 
However, multiband observations of NGC 4993 seems to be inconsistent with the presence of a globular 
or young star cluster with a total mass larger than a few thousands solar 
masses~\citep{gw170817noglobclust}. 
Nonetheless, a star cluster dominated by old faint and/or compact stars as the local environment 
of GW/GRB 170817A event may evade this constraint. For instance, the absolute magnitude limit of 
$M_V > -6.7$ concluded for any star cluster from observations of~\citep{gw170817noglobclust} is much 
brighter than the old faint star cluster Segue 3~\citep{segue3globclust,segue3globclust0} which has 
an absolute magnitude of $M \sim 0$ or even a star cluster $\sim 400$ fold brighter than Segue 3. 
In addition, a star cluster consisting of an old population as environment of GW/GRB 170817A event 
is consistent with conclusions obtained from analysis of its prompt gamma-ray emission 
in~\citep{hourigw170817}.

Alternatively, the late afterglow may be the result of multiple collisions between density shells 
in a mildly relativistic outflow rather than one collision. In this case emissions from shocks 
occurring closer to the center must pass through the outflow and will be subject to absorption. 
This scenario is similar to internal shocks in a relativistic jet, but here the flow is only mildly 
relativistic and shells will need much more time to catch with each other and collide. Our 
simulations show that in such a model shocks should have properties similar to model No. 3 if only 
a few shell collisions occurred or model No. 5 if many collisions had taken place. The column 
densities of these models are consistent with the estimation of $N_H$ in short 
GRBs~\citep{grbshortenv}. It is also enough to provide the necessary extinction.

In absence of optical/IR extinction an excess of X-ray emission, most probably from kilonova 
remnant, is necessary to explain the observations. A thermal emission in X-ray from a very hot 
ejected disk at $\gtrsim T+100$ seems unlikely. In fact observation of the counterpart in far IR 
at $\sim T+264$~days shows that at this epoch the kilonova is cooled and its dominant emission 
is in far IR~\citep{gw170817lateir}. On the other hand, X-ray from decay of nuclides produced in 
the kilonova and/or recombination of electrons in the ejected disk are credible sources of addition 
X-ray. Indeed, hundreds of electronic excitation lines and isotopes with half-life of 
$\mathcal{O}(100)$ days exist, including those related to nuclides produced through 
r-processes~\cite{nucldatabase}. 

This alternative solution needs a quantitative investigation, which must include calculation of 
isotopes yield and their nuclear excitation state in the kilonova and evolution of their ionization 
state. This task is out of the scope of the present work. Nonetheless, late time X-ray luminosity of 
GW/GRB 170817A afterglow of $\sim \mathcal{O}(1) \times 10^{39}$~\citep{gw170817latexary} is similar 
to those of supernovae type II~\citep{sntypeiixray}. Although the amount of ejected material from 
SN type II is much larger than in a BNS merger, most of the material is from the envelop of the 
star, which has a low level of nuclear excitation. By contrast, BNS ejecta comes from a much denser 
environment where QCD interaction prevails and includes freshly produced and highly excited isotopes. 
In addition, GW/GRB 170817A is one of the closest extra-galactic transient ever observed. For 
instance, it is 3 times closer than type II SN 1988Z which is extensively observed in 
X-ray~\citep{sntypeii1988z}. Therefore, radiation from decay of nuclides may have observable 
contribution.

\section{Outlines} \label{sec:outline}
In conclusion, shocks generated by the collision of a mildly relativistic outflow with 
a sheath of material surrounding the merger, or density layers inside the outflow can explain 
brightening and other characteristics of X-ray, optical/IR, and radio counterparts of 
GW/GRB 170817A. We suggested two scenarios for the generation of these afterglows. 

In the first scenario synchrotron self-absorption and extinction of optical emission in the 
environment of the source are necessary to explain late multi-band observations. The most probable 
origin of the external extinction, which may be additionally responsible for the faintness of 
undetected early afterglow of the GRB 170817A, is the presence of an old star cluster surrounding 
the progenitor BNS. This conclusion is consistent with larger probability of the formation of BNS 
and their merger in star clusters than in low density regions of galaxies, and conclusions 
of~\citep{hourigw170817} from the analysis of prompt gamma-ray emission. 

In the second scenario multiple shocks generated by collision of density shells inside the same 
outflow provide both the emission and absorption of optical emission with respect to X-ray that 
we find necessary for explaining the data. 

An excess of X-ray rather than extinction of optical emission, most probably from decay of nuclides 
and/or recombination of electrons in the kilonova remnant is another possibility which must be 
considered and quantified in order to confirm or refute this hypothesis.

\paragraph*{Note:} After completion of this work, GCN 23137 and GCN 23140 (Haggard, \etal), and 
GCN 23139 (Dobie, \etal) reported observations of afterglows of GW/GRB 170817A at 
$\gtrsim T + 300$~days in X-ray and radio bands, respectively. The observed sharp decline of 
radio emission at this time is consistent with Simul. No. 6 shown in Fig. \ref{fig:lc}. Steeper 
decline of the X-ray light curve and its absorption, which is implicitly an evidence for optical 
absorption as discussed above are consistent with predictions of this work. Additionally, 
Mooley \etal 2018 (arXiv:1806.09693) report the detection of superluminal motion of a relativistic 
jet/outflow in VLBI observations. Analysis of this data is consistent with a Lorentz factor of 
$\sim 3$ at $\sim T+150$ days, which is consistent with simulations 1, 6, and two of simulations 
shown in Fig. \ref {fig:n0}-b and c.

\paragraph*{Acknowledgment:} The author thanks the anonymous referee of~\citep{hourigw170817} for 
encouraging application of~\citep{hourigrb,hourigrbmag} model to late afterglows of GW/GRB 170817A.

\appendix
\section{Synchrotron self-absorption} \label{app:abs}
Reduction of photon flux of a source due to absorption during propagation of photons in matter is 
a Possinian process and can be formulated as:
\be
\frac{dI_\nu}{d\ell} = \alpha_\nu I_\nu    \label{abscoeffdef}
\ee
where $I_\nu$ is intensity at frequency (energy) $\nu$ and $\ell$ is propagation distance inside 
an absorbing material. For synchrotron self-absorption the absorption coefficient $\alpha_\nu$ 
can be related to distribution of accelerated electrons~\citep{emissionbook}:
\be
\alpha_\omega = \frac{\pi}{2\omega} \int_{\gamma_e}^\infty d\gamma_e P(\omega, \gamma_e) \gamma_e^2 
\frac{\partial}{\partial \gamma_e} \biggl (\frac{n'_e(\gamma_e)}{\gamma_e^2} \biggr ) 
\label {abscoefsynch}
\ee
where $\omega = E/\hbar$ is photon mode for a photon of energy $E$; $\gamma_e$ is Lorentz factor 
of accelerated electrons; $n'_e(\gamma_e)$ is number density of electrons with Lorentz factor 
$\gamma_e$; and $P(\omega,\gamma_e) \equiv dP/\omega d\omega$ is differential synchrotron power 
density for mode $\omega$. 

Using the phenomenological expression of $P(\omega,\gamma_e)$ obtained in~\citep{hourigrb}, the 
absorption coefficient $\alpha_\omega$ can be written as:
\be
\alpha_\omega = \frac{\sqrt{3} e^2}{\omega \gamma_m^2} \int_1^\infty d\eta F (\eta) 
\biggl [\int^\infty_{\omega / (\omega_m \eta)} d\zeta ~ K_{5/3} (\zeta) + \ldots \biggr ], \quad \quad 
F(\eta) \equiv \frac{\partial}{\partial \eta}\biggl (n'_e(\eta) \biggr ) \label {abscoefphenom}
\ee
where $K_\nu$ is the second modified Bessel function of order $\nu$ and dots mean higher order 
subdominant terms which depend on the geometry of the emitting surface. They are neglected in our 
simulations. This is a good approximation for the prompt emission, in which due to the large 
Lorentz factor of jet, the effective opening angle visible to a far observer is small and emission 
is highly beamed. For late afterglows the effect of high latitude emission can be significant and 
should be added to other uncertainties of the model. In particular, high latitude emission 
increases the total duration of afterglow for a given detection threshold. Nonetheless, even for 
a Lorentz factor as small as 2 the visible opening angle is $30^\circ$ and delay for arrival of 
photons from high latitudes would be $\sim 13\%$ of the time necessary for light to traverse a 
distance equal to the radius of emission surface. For an initial distance of 
$\mathcal{O}(1) \times 10^{16}$~cm used in our simulations and the final radius of $\sim 10$ times 
larger, the delay is $\lesssim 10$~days i.e. comparable with the exposure time in the latest 
observations. Thus, the increase in emission duration due to high latitude emission is of the same 
order as uncertainty on the observation time and does not have significant impact on the comparison 
of models with the data.

Equation (\ref{abscoefphenom}) shows that the synchrotron self-absorption coefficient 
depends on $\eta \equiv \gamma'_e / \gamma'_c$, 
$\omega'_c\equiv \frac {3e \gamma_e^2 B'_\bot}{2c m_e}$, and $\omega_m = \omega_c|_{\gamma_e = \gamma_m}$ 
where $\gamma_m$ is the minimum Lorentz factor of accelerated electrons. The second integral has 
an approximate analytical expression:
\be
\int^\infty_{\omega / (\omega_m \eta)} d\zeta ~ K_{5/3} (\zeta) \approx 2 K_{2/3} (\omega / (\omega_m \eta))
\label{besselintapprox}
\ee
We use this approximate expression in our simulations.

The total amount of synchrotron self-absorption is obtained from integrating 
eq. (\ref{abscoeffdef}). It depends on the geometric extension of magnetic field in the shock 
front $D_{sy}$. In our simulations we assume that this length is proportional to synchrotron 
characteristic time~\citep{emissionbook} and reduces exponentially or according to a 
power-law with propagation of shock front, i.e. $D_{sy} (r) = ct_{sy}(r) f(r/r_0)$ where 
$t_{sy} (r) = 3c\pi \gamma_m^3 (r)/\omega'_c (r)$ is the synchrotron characteristic time for 
electrons with minimum energy; $c$ is the speed of light. In the simulations we consider either 
$f(r/r_0) = \exp (-\alpha r/r_0)$ or $f(r/r_0) = (r/r_0)^\alpha$. 

\section{Evolution models of active region} \label{app:modes}
In the phenomenological model of~\citep{hourigrb} the evolution of $\Delta r'(r')$ cannot be 
determined from first principles. For this reason we consider the following phenomenological 
models:
\bea
&& \Delta r' = \Delta r'_0 \biggl (\frac {\gamma'_0 \beta'}{\beta'_0 \gamma'} 
\biggr )^{\tau}\Theta (r'-r'_0) \quad \text {dynamical model, Model = 0} \label {drdyn} \\
&& \Delta r' = \Delta r'_{\infty} \bigg [1-\biggl (\frac{r'}{r'_0} \biggr )^
{-\delta}\biggr ] \Theta (r'-r'_0) \quad \text {Steady state model, Model = 1} \label {drquasi} \\
&& \Delta r' = \Delta r'_0 \biggl (\frac{r'}{r'_0} \biggr )^{-\delta} 
\Theta (r'-r'_0) \quad \text {Power-law model, Model = 2} \label {drquasiend} \\
&& \Delta r' = \Delta r_{\infty} \bigg [1- \exp (- \frac{\delta(r'-r'_0)}{r'_0}) \biggr ] 
\Theta (r-r'_0) \quad \text {Exponential model, Model = 3} \label {expon} \\
&& \Delta r' = \Delta r'_0 \exp \biggl (-\delta\frac{r'}{r'_0} \biggr )
\Theta (r'-r'_0) \quad \text {Exponential decay model, Model = 4} \label {expodecay}
\eea
The initial width $\Delta r'(r'_0)$ in Model = 1 \& 3 is zero. Therefore, they are suitable for 
description of initial formation of an active region in internal or external shocks. Other models 
are suitable for describing more moderate growth or decline of the active region. In Table 
\ref{tab:param} the column $mod.$ indicates which evolution rule is used in a simulation regime - as 
defined in the foot notes of this table - using model number given in \ref{drdyn}-\ref{expodecay}.

\end{document}